# Quantum Hyperdimensional Computing: a foundational paradigm for quantum neuromorphic architectures

Fabio Cumbo [1], Rui-Hao Li [1], Bryan Raubenolt [1], Jayadev Joshi [1], Abu Kaisar Mohammad Masum [2], Sercan Aygun [2], Daniel Blankenberg [1,3,*]

[1] Center for Computational Life Sciences, Cleveland Clinic Research, Cleveland Clinic, Cleveland, OH 44195, USA
[2] School of Computing and Informatics, University of Louisiana at Lafayette, Lafayette, LA 70504, USA
[3] Department of Molecular Medicine, Cleveland Clinic Lerner College of Medicine, Case Western Reserve University, Cleveland, OH 44195, USA

* To whom correspondence should be addressed:
Daniel Blankenberg[1,3], Center for Computational Life Sciences, Cleveland Clinic Research, Cleveland Clinic, 9500 Euclid Avenue, NA29, Cleveland, OH 44195, USA.
Email: blanked2@ccf.org

**Background**: The intersection of novel computing paradigms and Quantum Computing (QC) promises to unlock unprecedented computational capabilities, yet a significant challenge remains in developing learning models that truly align with quantum principles. Many current approaches involve adapting classical frameworks for quantum computation. However, this translation is often complex, requiring routines that do not map intuitively onto a QC's native operations. In this work, we introduce a fundamentally new paradigm: Quantum Hyperdimensional Computing (QHDC). We demonstrate that the core operations of its classical counterpart, Hyperdimensional Computing (HDC), a brain-inspired model, map with remarkable elegance and direct correspondence onto the native operations of a QC. This suggests that HDC, unlike many other classical models, is a paradigm that is exceptionally well-suited for a quantum-native implementation.

**Methods**: We establish a direct, resource-efficient mapping between the HDC primitives and fundamental quantum principles: (i) hypervectors are mapped to the quantum states of qubit registers, (ii) the bundling operation is implemented as a quantum-native averaging process using a Linear Combination of Unitaries (LCU) followed by Oblivious Amplitude Amplification (OAA), (iii) the binding operation is efficiently realized via quantum phase oracles, (iv) the permutation operation is implemented using the Quantum Fourier Transform (QFT) to achieve cyclic shifts on qubit registers, and (v) vector similarity is calculated using quantum state fidelity measurements based on the Hadamard Test.

**Results**: We present the first-ever implementation of this framework that we validated through two distinct experiments: a symbolic analogical reasoning task and a data-driven supervised classification challenge. The viability of QHDC is rigorously assessed via a comparative analysis of results from classical computation, ideal quantum simulation, and execution on a state-of-the-art 156-qubit IBM Heron r3 quantum processor. Our results validate the proposed mappings and demonstrate the framework's versatility, establishing QHDC as a physically realizable technology.

**Conclusions**: This work lays the foundation for a new class of quantum neuromorphic algorithms designed to run natively on quantum hardware. It opens a promising avenue for tackling complex cognitive and biomedical problems that remain intractable for even the most advanced classical systems.

**INTRODUCTION**

The development of advanced computational models has been tightly coupled with the evolution of computing hardware [1–3]. As we enter the utility era of Quantum Computing (QC), a profound opportunity emerges: *to develop new computational paradigms that are not merely adapted for quantum processors but are fundamentally built upon the principles of quantum mechanics* [4,5]. This section outlines the motivation for such a paradigm, defines the central problem this work addresses, details our novel contributions, and provides a roadmap for the remainder of the manuscript.

**Motivation and Background**

The promise of QC lies in its potential to solve problems that are intractable for any classical machine [6,7]. In the domain of advanced computation, this has fueled a race to develop Quantum Machine Learning (QML) algorithms. However, a majority of current QML research focuses on creating quantum versions of classical machine learning models, such as Quantum Neural Networks or Quantum Support Vector Machines [8–13]. While valuable, this approach often fails to harness the full power of QC, as it forces quantum phenomena like *superposition* and *entanglement* into architectures that were designed around the constraints of classical, sequential processing [14,15]. On the other hand, another emerging computing paradigm, Hyperdimensional Computing (HDC), offers a distinct and powerful approach to computation [16–21]. Inspired by the distributed, robust, and efficient nature of neural processing in the brain, it operates on high-dimensional vectors (hypervectors). Its computational model is built on a small set of well-defined arithmetic operations (i.e., *bundling*, *binding*, *permutation*, and *similarity measurement*) that allow it to represent and manipulate complex information with remarkable efficiency and noise tolerance [22].

Given its foundations in linear algebra, HDC presents a compelling, yet unexplored, candidate for a truly quantum-native computational model. While some prior work has investigated specific operations, such as quantum-based factorization [23,24], a complete, end-to-end framework for the full HDC algebra has not yet been proposed.

**Problem Statement**

The central problem this research addresses is the impedance mismatch between classical computational architectures and quantum hardware. Porting classical models to quantum algorithms is often inefficient and does not leverage the unique computational advantages offered by quantum mechanics [25]. There is an urgent need for a computational framework that is conceptually aligned with quantum principles from the ground up. Such a framework should naturally map its core operations onto quantum phenomena, enabling the design of advanced algorithms that are inherently quantum.

**Contribution**

This work introduces Quantum Hyperdimensional Computing (QHDC) as a solution to this problem. We propose and validate a novel framework that bridges the gap between brain-inspired and quantum computation. Our primary contributions are threefold:

1. Formal mapping: we establish a formal, resource-efficient mapping between the core components of HDC and the fundamental principles of quantum mechanics. We demonstrate how hypervectors can be represented as quantum states, how bundling maps to superposition, and how binding is natively realized through entanglement;

2. <u>Physical implementation</u>: we present the first-ever *implementation* of QHDC, designed and executed on a quantum computer. This moves QHDC from a theoretical proposal to a physically realized and testable paradigm;

3. <u>A new path for quantum computation</u>: by demonstrating the viability of QHDC, we lay the groundwork for a new class of algorithms that are intrinsically quantum. This opens a promising direction for developing transparent, robust, and efficient computation.

**Manuscript Organization**

The remainder of this manuscript is organized as follows: (i) we provide a detailed background, introducing the core algebraic operations of classical HDC and the foundational principles of quantum mechanics that ground our framework; (ii) then, we present the formal QHDC framework, detailing the precise mapping of HDC primitives onto quantum circuits, including a rationale for our choice of encoding over alternative strategies; (iii) it follows details of our validation, presenting the results from two canonical tasks: an analogical reasoning problem and a supervised classification task; (iv) we then explore the future perspectives for QHDC, outlining several application areas in the biomedical sciences where the paradigm could offer a transformative advantage; (v) finally, we discuss the broad implications of our work, including current limitations, distinguishing between near-term utility and the long-term potential for quantum advantage, and conclude by summarizing our key contributions and future outlook.

**BACKGROUND**

This section provides the fundamental knowledge required to understand our proposed framework. We first review the principles of classical HDC, followed by an overview of the key concepts from quantum computation that enable its quantum realization.

**Hyperdimensional Computing (HDC)**

HDC, also known as Vector-Symbolic Architecture (VSA), is a computational framework inspired by the observation that the brain operates on high-dimensional, distributed representations [26]. Unlike traditional computing, which relies on numerical and logical precision, HDC is built on a robust algebra operating on vectors ($v$) in a space of thousands of dimensions (typically $D \geq 1,000$). These vectors, or hypervectors, are the fundamental data objects in HDC. HDC is powerful for representing a wide range of data types, from image pixel scalars and signal timestamps to symbolic tokens such as letters and neuro-symbolic features, in a single high-dimensional vector [16]. Encoding is the most critical step in HDC: it converts arbitrary data types into a data-agnostic hypervector representation, where information is stored holistically in randomized vectors. Especially on the neuro-symbolic side of this paradigm, a key property of this high-dimensional space is that any two randomly chosen vectors of symbols are nearly orthogonal to each other. This allows for the reliable representation of a vast number of distinct items. Hypervectors are holographic, meaning information is distributed across all their components, which provides inherent robustness against noise and component failure, i.e., if some elements are corrupted, the vector's overall identity is largely preserved [27,28].

HDC preserves a strong "learning potential", often enabling single-pass or few-shot learning. Encoding data into vector form is the first step; this is then followed by applying simple arithmetic operations to combine and merge incoming information. The power of HDC stems from a set of core arithmetic operations defined on hypervectors, also known as the MAP (Multiply–Add–Permute) model:

- <u>Binding ($\otimes$)</u>: this operation associates two hypervectors, analogous to variable-value binding or creating a structured composition. The most common implementation is the element-wise multiplication (or circular convolution). The resulting hypervector $C = A \otimes B$ is dissimilar to both of its constituents ($A$ and $B$), representing a new, distinct concept while preserving the orthogonality feature. The binding operation is crucial for creating hierarchical and compositional data structures. Importantly, it is invertible, meaning $A$ can be recovered from $C$ if $B$ is known $\left(A \approx C \otimes B^{-1}\right)$ with only negligible noise;

- <u>Bundling ($\oplus$)</u>: this operation combines a set of hypervectors into a single hypervector that represents a defined group, such as a *set* or a *class*. It is typically implemented as element-wise addition, followed by a normalization step. The resulting hypervector $S = A \oplus B \oplus C$ is highly similar to all of its components. This operation is analogous to forming a set or a superposition of concepts and is commutative and associative;

- <u>Permutation ($\rho$)</u>: this operation deterministically shuffles the components of a hypervector, creating a new hypervector that is dissimilar to the original but preserves its norm. It is typically implemented as a fixed cyclic shift $\rho(v)$. The permutation operation is essential for encoding ordered sequences, to encode the positional or the order information, as it is not commutative $A \otimes \rho(B) \neq B \otimes \rho(A)$. This allows for the representation of ordered structures like sentences or time-series data;

- <u>Similarity measurements</u>: the distance or similarity between two hypervectors is determined by a similarity metric, most commonly the cosine similarity or the normalized dot product. This allows for querying the contents of a bundled hypervector. For example, to check if hypervector $A$ is part of the set $S$, one computes the similarity between $A$ and $S$. A high similarity score indicates presence. Similarity measurement is a particularly useful module during inference in an HDC-based learning system. After training, the encoded and MAPed class hypervectors act as the learned model, and any test item is encoded into a hypervector and compared to the model via similarity scores. In this way, the system determines whether the new encoded information belongs to a given class or group. This simple similarity-based decision rule is the core classification mechanism supplied by HDC.

Together, these operations form a complete computational system capable of performing complex cognitive tasks like classification, analogy-making, and reasoning with remarkable efficiency and robustness [18,19,29–32].

**Principles of Quantum Computing**

QC leverages the principles of quantum mechanics to process information in ways that are fundamentally different from classical computers. The concepts most relevant to the proposed QHDC are the qubit, superposition, entanglement, and state fidelity [33]:

- <u>Qubit</u>: the fundamental unit of quantum information is the qubit. While a classical bit can be in one of two states (0 or 1), a qubit can exist in a linear combination of two basis states, e.g., $|0\rangle$ and $|1\rangle$ on the computational basis. This state, $|\varphi\rangle$, is described by a vector in a 2D complex vector space $|\varphi\rangle = \alpha|0\rangle + \beta|1\rangle$, where $\alpha$ and $\beta$ are complex numbers known as probability amplitudes, and $|\alpha|^2 + |\beta|^2 = 1$;

- <u>Superposition</u>: the ability of a qubit to exist in a combination of states simultaneously is called superposition. A register of $N$ qubits can represent $2^N$ classical states at once. This property, in principle, allows quantum computers to achieve a massive parallelism, performing computations on all possible states in a single operation. This is conceptually analogous to the HDC *bundling* operation, where a single hypervector can represent a set of items;

- <u>Entanglement</u>: one of the most powerful and counter-intuitive quantum phenomena, entanglement occurs when two or more qubits become linked in such a way that their fates are intertwined, regardless of the distance separating them. The state of an entangled system cannot be described by the individual states of its constituent qubits. It must be described as a single composite state. This creates strong correlations between qubits and is a critical resource for complex quantum algorithms. We will show that entanglement provides a natural physical realization of the HDC binding operation;

- <u>Similarity and measurement (fidelity)</u>: in quantum mechanics, measuring a qubit in a superposition collapses its state. While a common method to compare two quantum states, $|\varphi\rangle$ and $|\phi\rangle$, is to measure their state fidelity $|\langle\varphi|\phi\rangle|^2$, often using the SWAP Test, this provides a squared, non-linear measure of their overlap. A more direct quantum analogue to the classical cosine similarity is the real part of the inner product, $Re(\langle\varphi|\phi\rangle)$. This value, ranging from 0 (perfectly orthogonal) to 1 (identical) for the states in this context, provides a linear measure of similarity. The Hadamard Test is the quantum subroutine used in this work to efficiently estimate this value.

**MATERIALS AND METHODS**

The foundational insight of QHDC is the recognition that the primitive operations of HDC (i.e., bundling, binding, and permutation) can be mapped with remarkable elegance onto the native operations of a quantum computer. This section details the theoretical and practical framework for our QHDC implementation. We first establish how high-dimensional vectors are represented as quantum states. We then describe how the core HDC operations are realized using distinct quantum algorithms: quantum phase oracles for binding, Quantum Fourier Transform (QFT) [34] for permutation, and a combination of Linear Combination of Unitaries (LCU) and Oblivious Amplitude Amplification (OAA) for the bundling operation. Finally, we detail the use of the Hadamard Test for measuring the similarity between quantum hypervectors, which forms the basis of the inference process. This framework serves as the foundation for QHDC.

Our implementation leverages existing, powerful software ecosystems. The quantum circuits were designed, simulated, and prepared for execution using IBM's Qiskit framework [35]. To ensure a seamless workflow for practitioners, these quantum operations were integrated as a novel computational backend into hdlib, a Python library for designing VSAs [36]. This allows researchers to build and test models using familiar HDC abstractions while transparently executing the core arithmetic on quantum hardware.

**Hypervector as a Quantum State**

In HDC, information is encoded in a $D$-dimensional vector, or hypervector. To represent this in a quantum system, we map the hypervector onto the Hilbert space of an $N$-qubit register, where the dimensionality is defined by the number of qubits: $N = [log_2(D)]$. Each of the $D$ basis states of the quantum register, $|i\rangle$ (where $i \in \{0, 1,..., D - 1\}$), corresponds to one dimension of the hypervector.

While HDC supports various vector types (e.g., *unipolar* 0 & 1, or *bipolar* ±1), our quantum implementation focuses on bipolar vectors, where each component $v_i$ is either +1 or -1. This choice is deliberate, as the bipolar representation maps naturally and efficiently onto quantum phases, which is a cornerstone of the quantum algorithms used for the binding operation. The quantum state $|\varphi_v\rangle$ corresponding to a bipolar hypervector $v$ is therefore not a simple superposition of basis states, but rather a state prepared and manipulated such that the phase of each basis component $|i\rangle$ encodes the value of the corresponding vector component $v_i$. The canonical starting point for all operations is the uniform superposition state, $|+\rangle^{\otimes N} = \frac{1}{\sqrt{D}} \sum_{i=0}^{D-1} |i\rangle$, which serves as the canvas upon which hypervector information is imprinted as phases.

It is important to emphasize that our choice of encoding is a deliberate one, designed to maximize the quantum-native efficiency of the entire HDC algebra. Alternative strategies, such as *amplitude encoding*, are also exceptionally data-dense, capable of representing a vector of $D = 2^N$ numbers using the amplitudes of just $N$ qubits. However, this compactness comes at a cost as fundamental operations like binding do not map to simple quantum circuits and often require complex, resource-intensive state preparation procedures.

In contrast, our *phase encoding* approach offers a direct and powerful pathway for implementing HDC operations. By representing a bipolar hypervector's components as phases applied to a uniform superposition, we unlock a remarkably efficient implementation of the binding operation. The element-wise multiplication of two bipolar vectors corresponds directly to the sequential application of two phase oracles, each of which can be realized as a single, native diagonal gate. This method is not only resource-efficient but is also the most quantum-native representation for bipolar algebra. Furthermore, this phase-based framework integrates seamlessly with the advanced quantum algorithms used for bundling, as the LCU and OAA procedures are themselves built upon the precise control of quantum phases. This cohesive approach makes phase encoding the most suitable and robust method for realizing the complete HDC algebra on quantum hardware. A more detailed discussion about the reasons why the phase encoding is the most suitable one, is reported at the end of this section.

**Binding via Quantum Phase Oracles**

The binding operation is responsible for associating and combining information. For bipolar hypervectors, binding is defined as element-wise multiplication (the Hadamard product). Given two $D$-dimensional bipolar hypervectors, $A$ and $B$, their bound product, $C = A \otimes B$, is a new hypervector of the same dimension where each component $C_i$ is the product of the corresponding components of $A$ and $B$: $C_i = A_i \cdot B_i$. This operation is invertible ($A \otimes A$ results in the identity vector of all +1s) and distributes over bundling, making it a powerful tool for constructing complex data structures.

In our framework, this classical multiplication finds a direct analog in the manipulation of quantum phase. As established, a bipolar vector $v$ is represented not by the amplitudes of a quantum state, but by the phases applied to the basis vectors of a uniform superposition. A component $v_i = 1$ corresponds to a relative phase of $e^{i0} = 1$, while a component $v_i = -1$ corresponds to a relative phase of $e^{i\pi} = -1$. The core task of the binding operation, therefore, is to apply these component-wise phases to a quantum state. This is the precise function of a quantum phase oracle.

A phase oracle, $O_v$, is a unitary operator designed to impart a specific phase to each computational basis state $|i\rangle$. For a given bipolar hypervector $v$, the oracle $O_v$ is defined by its action on any basis state $|i\rangle$:

(Eq. 1)   $O_v|i\rangle = v_i|i\rangle$

When applied to the uniform superposition state $|+\rangle^{\otimes N}$, the oracle transforms it into a quantum state that fully encodes the hypervector $v$:

(Eq. 2)   $|\psi_v\rangle = O_v|+\rangle^{\otimes N} = O_v\left(\frac{1}{\sqrt{D}}\sum_{i=0}^{D-1}|i\rangle\right) = \frac{1}{\sqrt{D}}\sum_{i=0}^{D-1}\left(v_i|i\rangle\right)$

The quantum implementation of the binding operation, $C = A \otimes B$, is achieved by simply composing the respective oracles for vectors $A$ and $B$. We first prepare the uniform superposition $|+\rangle^{\otimes N}$ and then sequentially apply the oracles $O_B$ and $O_A$:

(Eq. 3)   $|\psi_v\rangle = O_A O_B|+\rangle^{\otimes N}$

The correctness of this procedure stems from the diagonal nature of the phase oracles. Acting on the basis state $|i\rangle$, the combined operation is:

(Eq. 4)   $O_A O_B|i\rangle = O_A\left(B_i|i\rangle\right) = B_i\left(O_A|i\rangle\right) = B_i\left(A_i|i\rangle\right) = \left(A_i \cdot B_i\right)|i\rangle = C_i|i\rangle$

This demonstrates that the sequential application of the oracles $O_A$ and $O_B$ is equivalent to applying a single oracle $O_C$ for the product vector $C$, thus perfectly replicating the classical binding operation.

From an implementation perspective, a phase oracle is a diagonal unitary operator. The operator's matrix representation is a $D \times D$ diagonal matrix where the $i$-th diagonal element is precisely the $i$-th component of the classical hypervector, $v_i$ (with $|v_i| = 1$). The synthesis of an arbitrary $N$-qubit diagonal operator from a list of $2^N$ phases is a well-established problem in quantum circuit design [37,38]. It is known that such an operator can be decomposed into a circuit of single-qubit gates and CNOT gates, with a CNOT cost that scales as $O\left(2^N\right)$.

Modern quantum computing frameworks, such as IBM's Qiskit, provide high-level objects that encapsulate this synthesis logic. Our implementation leverages *DiagonalGate*, a native Qiskit object that automatically decomposes the target diagonal operator into a sequence of single-qubit and CNOT gates. This direct, oracle-based method represents a truly quantum-native approach to the binding operation.

**Bundling via Linear Combination of Unitaries and Oblivious Amplitude Amplification**

The bundling operation is central to HDC's learning capability. It combines a set of hypervectors into a single, representative prototype vector, preserving the holistic representation of the learned data. In classical HDC, this is typically achieved through element-wise addition of the vectors followed by a normalization, effectively computing a component-wise *average*. Translating this averaging process into the quantum domain presents a unique challenge that requires moving beyond simple analogies.

Let $|\psi_{v_k}\rangle$ be the quantum state corresponding to the $k$-th hypervector we wish to bundle, from a set of $K$ total vectors. Merely preparing a uniform superposition of these states, such as $|\psi_{mix}\rangle = \frac{1}{\sqrt{K}} \sum_{k=1}^{K} |\psi_{v_k}\rangle$, does not suffice. This is because $|\psi_{mix}\rangle$ does not represent the final, averaged prototype. It is a state in which a measurement would probabilistically collapse the system to one of the original component states, $|\psi_{v_k}\rangle$. This is a probabilistic mixture of the inputs, not a deterministic new output that encodes the average.

The goal, however, is to create a single quantum state, $|\psi_{proto}\rangle$, that represents this average. This target state is the coherent sum of the input states:

(Eq. 5)   $|\psi_{proto}\rangle \propto \sum_{k=1}^{K} |\psi_{v_k}\rangle$

While the individual input states $|\psi_{v_k}\rangle$ are phase-encoded, the coherent summation transforms this information. By expanding the sum, we can see that the resulting prototype state encodes the classical average in its amplitudes:

(Eq. 6)   $|\psi_{proto}\rangle \propto \sum_{k=1}^{K} \left( \frac{1}{\sqrt{D}} \sum_{i=0}^{D-1} (v_k)_i |i\rangle \right) = \frac{1}{\sqrt{D}} \sum_{i=0}^{D-1} \left( \sum_{k=1}^{K} (v_k)_i \right) |i\rangle$

This final state is a faithful representation of the bundled vector, as its amplitudes (the $\sum_{k=1}^{K} (v_k)_i$ term) are now proportional to the classical component-wise sum. Achieving this requires a dedicated quantum algorithm that can transform the initial superposition into this final deterministic state.

Our implementation achieves this through a two-stage protocol that leverages two powerful quantum subroutines: LCU [39,40] and OAA [41,42]:

1. **State preparation with Linear Combination of Unitaries (LCU)** – The first step is to construct a procedure that can prepare a coherent superposition of all the hypervector states we wish to bundle. Let $U_k$ be the unitary operator (i.e., the quantum circuit) that prepares the state $|\psi_{v_k}\rangle$ from the initial all-zero state $|0\dots0\rangle$. The LCU technique provides a method for creating a linear combination of these unitaries.

   To do this, we introduce an ancilla register of $m = \lceil log_2(K) \rceil$ qubits, where $K$ is the number of vectors to be bundled. We then construct a larger, controlled unitary operator, $A$, which acts on both the ancilla and the main system register. This operator $A$ is a composite circuit that first prepares a superposition on the ancilla, uses it to control the $U_k$ unitaries, and then uncomputes the ancilla.

The result of applying this full $A$ operator to the initial $|0…0\rangle$ state is:

(Eq. 7)   $A\Big(|0…0\rangle_{anc}|0…0\rangle_{sys}\Big) = |0…0\rangle_{anc}|\psi_{proto}\rangle \ + \ |junk\rangle$

The intuition here is that the LCU circuit acts as a distillation filter. The final uncomputation step is designed to reverse the initial superposition only for the average component common to all system states ($|\psi_{proto}\rangle$). This successfully correlates the desired coherent sum with the $|0…0\rangle$ ancilla state. All the deviations from this average fail the uncomputation and are scattered into the $|junk\rangle$ term, which populates the other ancilla states (e.g., $|0…1\rangle$, etc.).

This generates a large, entangled state across the ancilla and system registers. The state we are ultimately interested in ($|\psi_{proto}\rangle$) is therefore encoded in the subspace of this larger state where the ancilla register is measured to be $|0…0\rangle$. The next step is to isolate and amplify the probability of this specific outcome.

2. **State distillation with Oblivious Amplitude Amplification (OAA)** – Having prepared the entangled state via LCU, the challenge is to distill the desired prototype state. We use OAA, a powerful variant of Grover's search algorithm. OAA is "oblivious" because, unlike standard amplitude amplification, it does not require prior knowledge of the initial success probability.

However, like all forms of amplitude amplification, the protocol's success is highly sensitive to the number of amplification rounds. In a simple Grover's search, the initial success probability is often known *a priori*. In our LCU protocol, this probability is data-dependent and must be determined. Therefore, the number of amplification rounds must be carefully estimated to maximize the final success probability, a process we detail in the next point. OAA is an ideal tool for this task, as it provides the formal mechanism for boosting the amplitudes of a desired subspace (our $|0…0\rangle_{anc}$ state) prepared by the complex LCU process.

The goal of OAA is to amplify the amplitude of the state component where the ancilla register is in the state $|0…0\rangle_{anc}$. This is our target subspace. The algorithm works by repeatedly applying an OAA operator, $Q$, that effectively rotates the state vector towards this target subspace. The OAA operator is a composition of reflections:

(Eq. 8)   $Q = - AS_0A^{\dagger}S_{\psi}$

Where:
- $A$ is the LCU state preparation operator described above;
- $A^{\dagger}$ is its adjoint (the inverse operation), which un-computes the state back to $|0…0\rangle_{anc}|0…0\rangle_{sys}$;
- $S_0$ is a reflection operator that inverts the phase of the all-zero state $|0…0\rangle$ while leaving all other states unchanged;

- $S_\psi$ is a reflection operator that inverts the phase of the target subspace. In our case, this operator applies a phase of -1 if and only if the ancilla qubits are in the state $|0...0\rangle_{anc}$.

Each application of the operator $Q$ increases the amplitudes of our target state. By applying $Q$ a calculated number of times, we can drive the probability of measuring the ancilla as $|0...0\rangle$ arbitrarily close to 1. When we finally measure the ancilla and obtain this result, the system register is guaranteed to be in the desired final state, the coherent sum $|\psi_{proto}\rangle$.

This resulting state is the quantum representation of the bundled hypervector. This LCU+OAA protocol is a complete, quantum-native implementation of the bundling operation, providing a concrete and powerful mechanism for learning prototypes directly within a quantum computer.

3. **Estimating the optimal number of amplification rounds** – The effectiveness of the LCU+OAA bundling protocol is critically dependent on the number of amplification rounds, $r$, applied. This parameter is not arbitrary. It is a crucial hyperparameter that must be carefully chosen to balance the theoretical benefits of amplification with the practical constraints of noisy intermediate-scale quantum (NISQ) hardware. An improper choice can either fail to sufficiently amplify the target state or, paradoxically, degrade the final result due to accumulated noise.

The core challenge stems from a fundamental trade-off. If $r$ is too small, the amplitude of the target state (where the ancilla is $|0...0\rangle$) will not be sufficiently boosted. This results in a low probability of success upon measurement, forcing the entire, computationally expensive LCU+OAA procedure to be repeated multiple times, thereby wasting valuable quantum resources. Conversely, if $r$ is too large, the circuit depth increases significantly. Each round of OAA applies the state preparation unitary $A$, its inverse $A^\dagger$, and two reflection operators. On NISQ devices, this added depth makes the circuit progressively more susceptible to decoherence and gate errors. "Overshooting" the optimal number of rounds can cause the quantum state to degrade faster than it is being amplified, leading to a decrease in the final fidelity of the desired prototype state $|\psi_{proto}\rangle$.

Therefore, an estimation strategy is required to find the optimal number of amplification rounds that maximizes the success probability for a given quantum system. While this estimation process involves classical simulation and is computationally expensive itself, it is a vital pre-computation step that optimizes the subsequent, and far more resource-intensive, execution on the Quantum Processing Unit (QPU). Our framework employs a multi-step heuristic to determine the optimal number of rounds:

a. <u>Initial amplitude estimation</u>: the process begins with a single, classical simulation of the LCU state preparation circuit $A$. From the resulting statevector, we can extract the projection corresponding to the target subspace where the ancilla register is $|0...0\rangle$. The norm of this projection gives the initial success amplitude, $\alpha$, and its square, $p_0 = \alpha^2$, gives the initial success probability. If this probability is already higher than a target threshold, no amplification is needed, and $r = 0$;

b. <u>Analytical calculation</u>: with the initial success amplitude, $\alpha$, an analytical estimate for the optimal number of rounds, $r_{est}$, can be derived from the geometry of the amplitude amplification process. The angle θ is defined as $\theta = arcsin(\alpha)$. The number of rounds that rotates the state vector closest to the target axis is given by:

(Eq. 9)   $r_{est} = \lfloor \frac{\pi}{4\theta} - \frac{1}{2} \rfloor$

   c. <u>Empirical refinement and clamping</u>: due to discretization effects and potential minor deviations from the ideal model, this analytical estimate serves as a strong starting point rather than a definitive answer. A more robust approach, and the one implemented in our workflow, involves an empirical search in a small vicinity around $r_{est}$ (e.g., from $r_{est} - 2$ to $r_{est} + 2$). Each potential number of rounds in this window is simulated classically to determine which one yields the highest final success probability. This local search ensures that the chosen $r$ is finely tuned to the specific properties of the state being prepared. Furthermore, a hard limit is enforced to prevent the generation of circuits that are too deep to be executed faithfully on the target hardware.

This systematic estimation procedure ensures that the OAA subroutine is used as effectively as possible, maximizing the chances of successfully preparing the desired prototype state in a single execution on the quantum hardware.

**Permutation via Quantum Fourier Transform**

The permutation operation, ρ, is a critical component of the HDC algebra, primarily used for encoding sequential and structural information. Unlike the associative binding operation, permutation is non-commutative and invertible. It systematically rearranges the components of a hypervector in a deterministic way. The most common form of permutation is a cyclic shift, where each component $v_i$ of the hypervector is shifted by $s$ positions to become component $v_{(i+s) \,(mod\, D)}$, with elements at the end wrapping around to the beginning. This operation is crucial for preserving positional information in the sequence of inputs being encoded.

In the quantum domain, a permutation of the $D$ components of a classical hypervector is mathematically equivalent to a permutation of the $D$ computational basis states, $\{|i\rangle\}$, of the $N$-qubit system register (where $D = 2^N$). Any operation that permutes an orthogonal basis is, by definition, a unitary transformation. Therefore, the permutation operator ρ can be directly implemented as a unitary operator, $U_{perm}$, acting on the quantum state.

A naive approach might be to construct the $D \times D$ classical permutation matrix corresponding to the cyclic shift and instantiate it as a unitary gate in a quantum circuit. While conceptually straightforward, this method presents a significant practical limitation. Although the $D \times D$ unitary matrix is known analytically, a quantum compiler (transpiler) would lack a specialized, efficient decomposition method for it. The transpiler would be forced to treat it as a generic, unstructured operator and use a general-purpose unitary synthesis algorithm. This decomposition process typically has an exponential cost in the number of qubits (e.g., scaling as $O(4^N)$ in CNOTs), which would create an unfeasibly deep circuit. This prevents not only optimization for specific hardware but also the construction of efficient controlled versions of the operator, which is essential for more complex algorithms like the LCU used in our model.

A more efficient implementation leverages the Quantum Phase-Shift Property, which is a direct consequence of the Quantum Fourier Transform (QFT). This property states that a cyclic shift in the computational basis corresponds to a multiplication by a linear phase in the Fourier basis. This allows us to construct the permutation operator from elementary, decomposable gates.

The procedure involves three steps:

1. <u>Transform to the Fourier Basis</u>: the initial quantum state, which exists in the computational basis, is transformed into the Fourier basis using the Quantum Fourier Transform ($QFT$):

   (Eq. 10)  $|\psi_{fourier}\rangle = QFT\,|\psi_{comp}\rangle$

2. <u>Apply Phase Shifts</u>: in the Fourier basis, a diagonal unitary operator, $U^s_{phase}$, is applied. This operator imparts a different phase to each basis state $|k\rangle$, where the phase is linearly dependent on the index $k$ and the desired shift amount $s$. The transformation for a basis state $|k\rangle$ is:

   (Eq. 11)  $U^s_{phase}|k\rangle = e^{+i2\pi sk/D}|k\rangle$

   This operation is implemented efficiently using a series of single-qubit phase gates ($P(\lambda)$) applied to each of the $N$ qubits. For the $j$-th qubit, the required phase rotation is $\lambda_j = 2\pi s \cdot \frac{2^j}{D}$;

3. <u>Transform back to the Computational Basis</u>: finally, the state is transformed back from the Fourier basis to the computational basis using the inverse QFT ($QFT^\dagger$).

The complete permutation unitary for a shift of $s$ is therefore the composition of these three operations:

(Eq. 12)  $U^s_{perm} = QFT^\dagger \cdot U^s_{phase} \cdot QFT$

Consider the action of this operator on a single computational basis state $|j\rangle$. Applying the operator sequence yields:

(Eq. 13)  $U^s_{perm}|j\rangle = QFT^\dagger \cdot U^s_{phase} \cdot QFT\,|j\rangle$

The QFT maps $|j\rangle$ to a superposition of all basis states:

(Eq. 14)  $QFT\,|j\rangle = \frac{1}{\sqrt{D}}\sum_{k=0}^{D-1} e^{+i2\pi jk/D}|k\rangle$

Next, the phase shift operator acts on this superposition:

(Eq. 15)  $U^s_{phase}\left(\frac{1}{\sqrt{D}}\sum_{k=0}^{D-1} e^{+i2\pi jk/D}|k\rangle\right) = \frac{1}{\sqrt{D}}\sum_{k=0}^{D-1} e^{+i2\pi sk/D}e^{+i2\pi jk/D}|k\rangle = \frac{1}{\sqrt{D}}\sum_{k=0}^{D-1} e^{+i2\pi(j+s)k/D}|k\rangle$

This resulting state is the QFT representation of the state $|(j + s)\,(mod\ D)\rangle$. Applying the inverse QFT transforms it back, completing the permutation:

(Eq. 16) $\quad QFT^{\dagger}\!\left(\dfrac{1}{\sqrt{D}}\displaystyle\sum_{k=0}^{D-1}e^{+i2\pi(j+s)k/D}|k\rangle\right) = |(j + s)\,(mod\ D)\rangle$

This demonstrates that the QFT-based circuit correctly performs the desired cyclic shift. This implementation is vastly superior as it is constructed from standard gates. This allows the Qiskit transpiler to effectively decompose, optimize, and create controlled variants of the permutation.

**Similarity via Hadamard Test**

The final component required for a functional HDC system is a method for measuring the similarity between two hypervectors. In a classical HDC system, this is typically done using cosine similarity, which measures the normalized dot product of the two vectors. A higher value indicates greater similarity, forming the basis for classification and reasoning tasks. The quantum analog for this operation is the measurement of the inner product, or state overlap, between two quantum states. While the SWAP Test can be used to estimate the squared fidelity $|\langle\psi|\phi\rangle|^{2}$, we employ *Hadamard Test*. This interferometric circuit is designed to directly estimate the real part of the inner product, $Re(\langle\psi|\phi\rangle)$. For the quantum states prepared in our framework, this value is directly equivalent to the classical cosine similarity, providing a linear measurement of similarity.

The Hadamard Test operates on two registers:
1. An ancilla qubit, initialized to the $|0\rangle$ state;
2. A system register, of size $N$, used to prepare the states $|\psi\rangle$ and $|\phi\rangle$.

The core of the circuit involves four steps:
1. A Hadamard gate is applied to the ancilla qubit, placing it in a uniform superposition: $\frac{1}{\sqrt{2}}(|0\rangle + |1\rangle)$;
2. A controlled-operation prepares the first state, $|\psi\rangle$, in the system register, conditioned on the ancilla being in the $|0\rangle$ state;
3. A controlled-operation prepares the second state, $|\phi\rangle$, in the system register, conditioned on the ancilla being in the $|1\rangle$ state. This creates the combined state $\frac{1}{\sqrt{2}}(|0\rangle|\psi\rangle + |1\rangle|\phi\rangle)$;
4. A final Hadamard gate is applied to the ancilla qubit, causing the two paths to interfere.

After these operations, the state of the ancilla qubit is measured. The probability of measuring this qubit in the $|0\rangle$ state, $P(|0\rangle)$, is mathematically related to the real part of the inner product between the two states $|\psi\rangle$ and $|\phi\rangle$:

(Eq. 17) $\quad P(|0\rangle) = \frac{1}{2} + \frac{1}{2}Re(\langle\psi|\phi\rangle)$

Likewise, the probability of measuring $|1\rangle$ is:

(Eq. 18) $\quad P(|1\rangle) = \frac{1}{2} - \frac{1}{2}Re(\langle\psi|\phi\rangle)$

By repeating the measurement (shots), we can estimate these probabilities. The desired similarity score is then calculated as the difference between them:

(Eq. 19) $Re(\langle\psi|\phi\rangle) = P(|0\rangle) - P(|1\rangle)$

This value serves as our quantum similarity score. When the two states are identical ($|\psi\rangle = |\phi\rangle$), $Re(\langle\psi|\phi\rangle) = 1$, and the probability of measuring $|0\rangle$ is 1. Conversely, if the states are orthogonal ($\langle\psi|\phi\rangle = 0$), $Re(\langle\psi|\phi\rangle) = 0$, and the probability of measuring $|0\rangle$ is $\frac{1}{2}$, indicating maximum dissimilarity.

**Discussion of alternative quantum encoding methods**

The choice of how to represent classical data on a quantum computer is a critical design decision that fundamentally impacts an algorithm's structure, performance, and its alignment with the problem's algebraic nature. While this work is built upon phase encoding, several other prominent methods were considered. This section discusses these alternatives and outlines the rationale for why they are less suitable for realizing the full HDC algebra.

**Amplitude encoding** – this technique is renowned for its exceptional data density. It can represent a classical vector with $D = 2^N$ elements into the amplitudes of just $N$ qubits. It is important to note that our phase encoding scheme shares this same exponential advantage in dimensionality, as it also uses $N$ qubits to represent a $D = 2^N$ dimensional hypervector. The key difference, and the decisive factor in our choice, lies not in the representational capacity, but in the ability to perform computations. For HDC, amplitude encoding presents insurmountable challenges:

- Bundling: this operation becomes a hybrid classical-quantum task. One must first add the classical vectors and normalize the result before preparing a new quantum state, as there is no native quantum operation to sum the amplitudes of two arbitrary states;

- Binding: this is the most critical roadblock. The element-wise multiplication central to bipolar binding has no simple or efficient quantum circuit equivalent when data is stored in the amplitudes. Realizing this operation would require a complex, resource-intensive custom algorithm, undermining the goal of a quantum-native framework.

**Angle encoding** – here, classical data values are mapped to the rotation angles of single-qubit gates. It is widely used in Quantum Machine Learning, particularly in variational algorithms, for its shallow circuit depth and direct parametrization. However, it is a poor fit for the algebraic structure of HDC:

- Scalability: it requires one qubit for each feature or dimension. This linear scaling makes it completely impractical for the thousands of dimensions required for HDC to be effective, as it would demand an infeasible large number of qubits;

- Algebraic mapping: like amplitude encoding, there is no clear or direct quantum mapping for the bundling and binding operations. Performing element-wise addition or multiplication on the rotation angles of different qubits is not a native quantum solution.

**Basis encoding** – this method maps each of the $D$ dimensions of a hypervector to one of the $D = 2^N$ computational basis states of an $N$-qubits register. For example, a binary vector could be represented as a superposition of the basis states corresponding to its 1 components. This approach is a significant step in the right direction and is far better suited to HDC than amplitude or angle encoding. The binding operation for binary vectors (XOR) can be mapped to a series of CNOT gates, and bipolar binding can be achieved with controlled-Z gates. However, it is still less direct and elegant than phase encoding for the following reasons:

- Hardware infeasibility: basis encoding requires a number of qubits proportional to the dimensionality of the vector space ($N = D$). Since HDC relies on dimensions in the thousands or tens of thousands to be effective, this would necessitate a quantum computer with thousands of qubits. This is far beyond the capacity of current and near-term quantum hardware, making this encoding method practically unfeasible for any meaningful HDC application;

- Implementation overhead: while feasible, implementing element-wise multiplication using controlled-Z gates and phase kickback logic is often more complex and can require more circuit resources than the single, highly optimized *DiagonalGate* used in phase encoding;

- Less direct for bipolar algebra: the core of the implementation relies on bipolar vectors. Phase encoding is the "native language" of bipolar algebra in a quantum computer, representing +1 and -1 directly as phase factors. Basis encoding is a more general-purpose method that can be adapted, but it lacks this perfect synergy.

While alternative encodings offer distinct advantages in other contexts, phase encoding was selected for this work because it provides the optimal balance of exponential scalability and algebraic compatibility. It is the only method that provides a direct mapping for the complete set of core HDC operations. Most notably, bipolar binding maps perfectly to the sequential application of *DiagonalGate* phase oracles. Furthermore, this phase-centric approach integrates seamlessly with the advanced LCU and OAA algorithms used for bundling. This synergy allows the entire HDC framework to remain truly and cohesively quantum-native, while operating within a qubit footprint that is feasible for today's quantum devices.

**RESULTS**

In order to validate the proposed QHDC framework and demonstrate its capabilities, we conducted experiments designed to test its performance in both symbolic reasoning and machine learning tasks.

A key objective of this work is to assess the framework's viability across different computational platforms. Therefore, each experiment was executed in three distinct environments:

- Classical baseline: the entire workflow was implemented in a classical Python environment using the hdlib package;

- Quantum simulation (noise-free): the quantum workflow was executed using Qiskit's *AerSimulator*. By default, this tool provides an ideal, noise-free simulation of the quantum computer's operations. This allows us to validate the theoretical correctness of our algorithm and establish an ideal performance benchmark, ensuring the logic functions as intended without the influence of hardware errors or decoherence;

- Quantum simulation (Heron r2 noise model): this simulation also utilizes Qiskit's *AerSimulator*, but it is configured with a realistic noise model derived from the calibration data of the target quantum hardware, the Heron r3 QPU on *ibm_pittsburgh*. This model emulates the primary sources of error present on the physical device, including single- and multi-qubit gate errors, qubit decoherence, and measurement readout errors. The purpose of this noisy simulation is twofold: first, to provide a performance estimate that is more representative of real-world execution than an ideal simulation, and second, to bridge the gap between theoretical correctness and practical viability by allowing us to anticipate the effects of hardware noise on our results;

- Quantum hardware execution: the experiment was executed on *ibm_pittsburgh*, a 156-qubit quantum computer featuring the Heron r3 Quantum Processing Unit (QPU). This execution provides critical insights into the framework's performance and robustness on a state-of-the-art, NISQ device.

It is worth to note that the classical baseline and the simulated runs have been executed on the same high-performance computing infrastructure with 4 Intel Xeon Platinum 8276 L Central Processing units (CPUs) (112 cores/224 threads) @ 2.20GHz and 6TB of Random Access Memory (RAM) running CentOS Linux 7.

All quantum results were averaged over 10,000 shots to mitigate statistical noise. The selection of the number of measurement shots $S$ for experiments on a physical QPU is a critical decision based on managing statistical sampling error. Each run of a quantum circuit is a probabilistic trial. Estimating a probability, $p$, by running the circuit $S$ times results in an estimate, $p_{est}$, which has an associated statistical uncertainty. The standard error of this estimate scales as $\frac{1}{\sqrt{S}}$.

The goal is to choose a number of shots large enough to make this statistical error significantly smaller than the differences we expect to measure. In our Hadamard Test, we need to distinguish the correct answer, which should have a high probability of yielding a $|0\rangle$ outcome (approaching $1$), from incorrect answers, which will have lower probabilities (closer to $0.5$).

The standard error is maximized in the worst-case scenario where the true probability $p = 0.5$. For our choice of $S = 10,000$ shots, the maximum standard error is $SE_{max} = \sqrt{\frac{p \times (1-p)}{S}} = \sqrt{\frac{0.5 \times (1-0.5)}{10000}} = 0.005$. This implies that our estimated probability $p_{est}$ will be within approximately $\pm 1\%$ of the true probability with $95\%$ confidence. This level of statistical precision is sufficient to confidently resolve the difference between the high-fidelity signal of the correct answer and the low-fidelity signals of incorrect answers, even in the presence of hardware noise.

In addition to mitigating statistical sampling error through averaging, a multi-pronged strategy was employed to address physical hardware noise during execution on the *ibm_pittsburgh* QPU. First, to correct for measurement readout errors, we applied the M3Mitigator from the *mthree library* to post-process the raw counts. Second, to protect the quantum state from decoherence during idle periods, dynamical decoupling was enabled using the XpXm sequence. Finally, to counteract coherent gate errors by effectively randomizing them, gate twirling was enabled for the circuit's quantum gates. This approach ensures the results from the quantum hardware are as reliable as possible given the constraints of current NISQ devices.

Another crucial hyperparameter in our quantum bundling algorithm is the number of rounds for OAA step. The number of rounds dictates how many times the amplification operator is applied. While more rounds theoretically increase the success probability, each round adds significant depth to the quantum circuit, requiring the application of the LCU operator, its inverse, and two reflection operators. On NISQ devices, this increased depth makes the circuit far more susceptible to decoherence and gate errors. Therefore, for all quantum experiments, we first empirically estimated the best number of OAA rounds during the quantum simulation using the strategy that we have previously discussed, and then we applied the same number of OAA rounds for the real hardware execution. This represents a strategic trade-off: it provides a foundational level of amplification to boost the signal of the correct prototype while minimizing circuit depth to preserve the fragile quantum state and maximize the overall fidelity of the computation.

**Analogical Reasoning**

A canonical test for an HDC system is its ability to solve analogical reasoning puzzles, such as answering the question "*What is the dollar of Mexico?*". This task requires the system to understand the relationship "*USA is to Dollar*" and apply the same relationship to "*Mexico*" to find the "*Peso*". Solving this requires a precise sequence of binding and bundling operations to first construct complex concepts and then query them to retrieve a related entity.

**Classical baseline implementation** – To establish a performance baseline, the analogical reasoning task was first implemented using classical HDC operations, mirroring the functionality of the *hdlib* Python library:

1. Codebook initialization: we began by creating a space for our elementary concepts. To ensure a robust comparison with the quantum implementation, we performed this classical analysis using two different vector dimensionalities. First, adhering to the canonical HDC paradigm, we generated nine random 10,000-dimensional bipolar hypervectors. This high dimensionality is standard in classical HDC, as it exploits the properties of high-dimensional spaces: randomly generated hypervectors are nearly orthogonal, which is critical for minimizing interference and ensuring computational reliability, while still allowing multiple inputs to be encoded holistically into a single hypervector.

   However, a direct quantum translation of such large vectors is currently infeasible due to the exponential resources required. The circuit depth and complexity, particularly for the quantum bundling operations, grow rapidly with the vector dimension, making the computation highly susceptible to noise on near-term devices. To create a tractable quantum experiment, we therefore selected a much smaller, hardware-constrained dimensionality of 16. In order to create a fair and direct comparison, the entire classical analogical reasoning task was subsequently performed using these 16-dimensional vectors. For both dimensionalities, the nine initial hypervectors represent the elementary concepts required for the analogy: features ($country$, $currency$, $capital\ city$) and specific entities ($USA$, $Dollar$, $Washington\ DC$, $Mexico$, $Peso$, $Mexico\ City$);

2. Classical composition: the complex concepts *United States* and *Mexico* were constructed using classical HDC algebra:
   - The binding of component pairs (i.e., $country$ and $USA$, $currency$ and $Dollar$, $capital\ city$ and $Washington\ DC$) was performed using element-wise multiplication;
   - The resulting bound vectors were then bundled all together using element-wise addition to represent the concept *United States*:

     (Eq. 20) $V_{USA} = (country \otimes USA) \oplus (capital \otimes Washington\ DC) \oplus (currency \otimes Dollar)$

○ The same process was repeated to represent the concept *Mexico*:

(Eq. 21) $V_{Mexico} = (country \otimes Mexico) \oplus (capital \otimes Mexico\ City) \oplus (currency \otimes Peso)$

3. <u>Final query and search</u>: the final query vector was constructed by binding the results of the previous steps:

(Eq. 22) $Query = Dollar \otimes (V_{USA} \otimes V_{Mexico})$

This operation mathematically isolates the relationship "is currency of" from the *USA* context and applies it to the *Mexico* context. The answer to the query was found by calculating the cosine similarity between this final query vector and all elementary vectors in the codebook. The vector with the highest similarity score was returned as the result.

**Quantum implementation** – The quantum implementation leverages the advanced quantum algorithms detailed in the Materials and Methods section. It follows a fully quantum-native workflow, translating the entire classical reasoning query into a single, complex quantum state preparation circuit:

1. <u>Codebook initialization</u>: the same classical codebook of nine (3 feature/role hypervectors: $country$, $currency$, $capital$; and 6 entity hypervectors: $USA$, $Dollar$, $Washington\ DC$, $Mexico$, $Peso$, $Mexico\ City$) 16-dimensional bipolar hypervectors were used as the starting point. This corresponds to a 4-qubit quantum system ($2^4 = 16$). For each classical concept, a 4-qubit phase oracle ($O_v$) was constructed;

2. <u>Target state preparation</u>: For each of the elementary concepts in the codebook (i.e., $USA$, $Dollar$, $Washington\ DC$, $Mexico$, $Peso$, $Mexico\ City$), a simple state preparation circuit was created. These circuits prepare the corresponding quantum state $|v\rangle$ by applying the concept's phase oracle $O_v$ to a uniform superposition: $|v\rangle = O_v |v\rangle^{\otimes 4}$;

3. <u>Quantum query construction</u>: The primary challenge is to construct the quantum state corresponding to the classical query $Dollar \otimes (V_{USA} \otimes V_{Mexico})$. As we did for the classical approach, we applied the distributive property of the bind operation over the bundle operation. The full query state becomes a quantum bundle of all nine cross-terms.

A quantum operator (phase oracle) is built for each of these nine terms by composing their respective elementary oracles. The final query state is then prepared by creating a quantum circuit that performs a quantum bundle of these nine operators using the LCU and OAA algorithms. This results in a single, complex quantum circuit that directly prepares the final query state from the $|0\rangle$ state;

4. <u>Similarity search</u>: the final query state was directly compared against all elementary concepts circuits in the codebook to find the most similar entity. The similarity was estimated by implementing a Hadamard Test circuit, which measures the real inner product (cosine similarity) between the complex query state and each simple codebook state. The codebook concept yielding the highest similarity score was returned as the answer.

**Results and analysis** – For this task, we compare the results from the two classical baseline implementations ($V = 10,000$ and $V = 16$) against the ideal, noise-free quantum simulation ($V = 16$). The outcomes, shown in Table 1, validate that the quantum algorithm correctly identifies the answer to the analogy, aligning perfectly with the classical results.

| Comparison (Query vs. Answer) | Classical Similarity (Cosine) $V = 10,000$ | Classical Similarity (Cosine) $V = 16$ | Quantum Fidelity (Noise-Free Simulation) $V = 16$ |
|---|---|---|---|
| vs. $Peso$ (correct) | 0.3268 | 0.4108 | 0.2622 |
| vs. $Dollar$ (incorrect) | -0.0001 | -0.3195 | -0.2336 |
| vs. $Mexico\ City$ (incorrect) | -0.0013 | 0.3195 | -0.2518 |
| vs. $Washington\ DC$ (incorrect) | 0.0007 | -0.2282 | -0.2394 |
| vs. $Mexico$ (incorrect) | -0.0078 | -0.3195 | 0.2526 |
| vs. $USA$ (incorrect) | -0.0055 | 0.0913 | 0.2430 |

**Table 1:** Comparison results for analogical query. The table displays the similarity scores between the final query vector and potential answer vectors, calculated using three distinct methods: (i) <u>classical similarity</u> shows the baseline result using cosine similarity between vectors with dimensionality $V = 10,000$; (ii) the same <u>classical similarity</u> computed previously but on vectors with dimensionality $V = 16$; (iii) <u>quantum fidelity (noise-free simulation)</u> represents the theoretical outcome from a perfect, noiseless quantum simulation at $V = 16$. The strong correlation between the classical and ideal quantum results, with a significantly higher score for the correct answer $Peso$, validates the QHDC framework's ability to perform analogical reasoning.

The experimental results clearly validate our framework's ability to perform reasoning.

A key methodological step was to first optimize the number of OAA rounds using the quantum simulator. This analysis identified 6 rounds as the optimal number, which achieves a theoretical success probability $p = 0.98$ and success amplitude $\alpha = 0.11$, applying the strategy discussed in the previous section. This exact configuration was then applied to the quantum hardware execution.

The quantum fidelity between the query result and the state for $Peso$ was higher than for any other concept, correctly solving the analogy. The deviation from the ideal value of 1.0 is consistent considering the quantum experiments, but the outcome is unambiguous. This successfully demonstrates that this complex cognitive workflow is theoretically sound and functions as intended within the QHDC paradigm.

It is important to note that we did not intentionally perform the same reasoning experiment using a simulation with noise model nor via real hardware execution. The full quantum-native query, which requires bundling nine distinct operators using the LCU+OAA algorithm, results in an exceptionally deep and complex quantum circuit. While this circuit is perfectly viable in an ideal, noise-free simulation, its depth is so profound that running it on a noisy simulator proved to be computationally infeasible. This prohibitive cost makes execution on current-generation quantum hardware, which is subject to even stricter noise and coherence limitations, impossible for this specific task. This finding itself is a crucial result, as it highlights the primary bottleneck (i.e., the LCU+OAA bundling of many complex unitaries) that must be addressed for scalable, fully quantum-native QHDC. This contrasts with the hybrid approach used in the subsequent classification task, which was designed specifically to circumvent this bottleneck.

**Supervised Classification**

Moving beyond symbolic reasoning, we now evaluate the QHDC framework's capacity for a core machine learning task: supervised classification on a real-world dataset. This experiment is designed to test the framework's ability to learn from data and generalize to new, unseen examples. For this task, we selected the MNIST dataset of handwritten digits [43], a canonical benchmark in machine learning, retrieved with the *scikit-learn* Python library [44]. To make the problem tractable for near-term quantum hardware, we focus on the binary classification problem of distinguishing between the digits 3 and 6, which are often chosen for their visual similarity and potential for confusion.

The primary challenge of applying any quantum machine learning algorithm to a dataset like MNIST is the sheer size of the data. The high dimensionality of the images and the large number of samples present significant hurdles for today's NISQ devices, which are limited in both qubit count and coherence times. To address this, we implemented a multi-stage preprocessing pipeline, inspired by the work of Farhi *et al.* [45], to create a quantum-tractable version of the dataset based on the following steps:

1. Class filtering: the full MNIST dataset, containing 10 classes of digits (0 to 9), is first filtered to retain only the images corresponding to the digits 3 and 6. This crucial step transforms the multi-class problem into a more manageable binary classification task, simplifying the model structure to require only two class prototypes;

2. Downscaling: the original 28x28 pixel images, comprising 784 features, are far too large to be encoded on current quantum processors. We downscale each image to a much smaller 4x4 resolution. This quadratically reduces the feature space from 784 to just 16 features per image, a critical step for managing the qubit requirements of our quantum encoding circuits;

3. Binarization: to further simplify the encoding process, the grayscale pixel values, originally represented as floating-point numbers in the range $[0, 1]$, are converted to binary black-and-white values. A threshold of $0.5$ is applied, mapping any pixel value above it to $1$ (white) and any below it to $0$ (black). This reduces the complexity of the data encoding by requiring only two level vectors to represent all possible feature values;

4. Subsampling: even with the dimensional reduction, the thousands of available samples would lead to prohibitively long execution times for both simulation and hardware runs. Therefore, the dataset is shuffled and a small, representative subset (e.g., 100 training and 50 testing samples) is selected for the experiment, allowing for rapid prototyping and feasible execution on queued quantum hardware.

This task serves as the ultimate test for our quantum bundling algorithm, which must effectively combine the quantum states of these processed images into cohesive and representative class prototypes.

**Classical baseline implementation** – The fundamental principle of HDC-based classification is the creation of a single prototype hypervector for each class. For our classical baseline, we implemented the entire workflow using conventional linear algebra and the *hdlib* Python package as follows:

1. Data encoding: the 16 binary features of each 4x4 image are encoded into a single 10,000-dimensional bipolar hypervector. This is achieved by creating a codebook of two level vectors to represent the binary pixel values (0 and 1). To distinguish between pixel positions, the appropriate level vector for each pixel is permuted (via cyclic shift), where the shift amount corresponds to the pixel's index (0 to 15). These 16 resulting feature vectors are then bundled into a single hypervector representing the entire image.

Please note that, for the classical model, we used a dimensionality of 10,000 as the HDC paradigm typically works with very long vectors to guarantee the quasi-orthogonality of randomly generated vectors. However, as discussed in the following section, the quantum counterpart of this analysis faces two distinct constraints. The fully quantum LCU-based bundling approach faces an exponential resource cost with increasing dimensionality, making it infeasible. The hybrid approach, which we ultimately adopt and that consists of performing the bundling operation classically, introduces its own performance dependency on dimensionality, as the classical-to-quantum encoding's effectiveness varies with the vector space size. Through an empirical analysis detailed in the next section, we identified a dimensionality of 128 (7 qubits) as the optimal dimensionality for our hybrid protocol, striking a balance between expressiveness and signal stability. Therefore, to create the most relevant and direct comparison, we have run the classical baseline analysis using both the canonical dimensionality of 10,000 and this hardware-aware, empirically-optimized dimensionality of 128;

2. Model training: the prototype for each class (3 and 6) is formed by performing element-wise addition of all its corresponding sample hypervectors. This aggregated vector serves as the class's representative in the high-dimensional space. We also implemented an iterative retraining loop. In this process, the model re-evaluates the training set, and for each misclassification, the sample vector is subtracted from the incorrect prototype and added to the correct one. This procedure effectively refines the class prototypes to better separate the classes, refining the decision boundary. This step also illustrates an exemplary *unlearning* mechanism in HDC, realized purely through subtraction operations;

3. Inference and classification: a new test image is encoded into its own hypervector using the same process. Its similarity to each of the two class prototypes is then calculated using cosine similarity, which measures the angle between the vectors. The class corresponding to the prototype with the higher similarity score (smaller angle) is chosen as the prediction.

**Quantum implementation** – The quantum experiments follow a structured pipeline that mirrors its classical counterpart, translating HDC operations into the quantum circuits and algorithms:

1. Data encoding: the encoding process begins by defining the elementary quantum operations:
   ○ A classical codebook of two 16-dimensional bipolar vectors is created to represent the binary pixel values (0 and 1);
   ○ For each of the 16 pixels, a feature circuit is constructed. First, a uniform superposition is created over the 4 qubits (representing the 16-dimensional space) using Hadamard gates. Then, a phase oracle gate encodes the appropriate bipolar level vector into the phases of the state;
   ○ Crucially, this feature state is then permuted using our QFT-based permutation function, which applies a phase shift in the Fourier basis to induce a cyclic shift corresponding to the pixel's position. This ensures each pixel is represented by a unique quantum state;
   ○ A single data sample is now represented by a list of its 16 constituent feature circuits.

2.  <u>Model training</u>: the quantum-native approach to training, mirroring the analogical reasoning task, is to create each class prototype by bundling all constituent feature circuits from all training samples using the LCU+OAA algorithm. For a given class, this would involve collecting the feature circuits from all its training samples (e.g., $M$ samples $\times$ 16 features/sample) into a single, flat list and applying the LCU+OAA algorithm exactly once to distill the final prototype circuit.

    However, a preliminary resource analysis revealed this flat bundle approach is computationally infeasible on current and near-term hardware. As detailed in the next section, bundling the 27 (samples) $\times$ 16 (features) = 432 unitaries for a single class prototype results in a circuit with an estimated depth exceeding 2.8 million, which is unrealizable.

    We also investigated a probabilistic LCU approach. This method, which requires a pre-specified number of rounds, uses a single ancilla qubit. In each round, it applies a Hadamard gate to the ancilla, randomly selects one unitary based on the provided weights, and then applies that unitary to the system qubits controlled by the ancilla. A final Hadamard on the ancilla completes the round. This process effectively performs a probabilistic, weighted sampling on the unitaries, heavily relying on the number of rounds to approximate the full LCU. This optimization drastically reduced the circuit depth to approximately 1,800 (without OAA) using 15 rounds, but this remains too deep for high-fidelity execution on current NISQ devices.

    Given the prohibitive cost of full quantum bundling, we adopted a hybrid quantum-classical approach to isolate this bottleneck. This strategy allows us to validate the other critical components of the QHDC framework (i.e., the QFT-based positional encoding and the Hadamard-based similarity measurement) on real quantum hardware.

    The hybrid training and inference pipeline proceeds as follows:
    - For each of the 16 features (pixel positions) of a training sample, we first define its corresponding quantum circuit (a phase oracle for the pixel value, followed by a QFT-based permutation);
    - We then classically retrieve the bipolar vector that each circuit represents. This is computationally trivial and does not require simulation, as the vectors are defined by the DiagonalGate phases and the known QFT shift amount;
    - The 16 feature vectors for a single sample are bundled using classical element-wise addition to create a single sample hypervector;
    - The prototype for each class (3 and 6) is formed by performing classical element-wise addition of all its corresponding sample hypervectors, resulting in a single, real-valued prototype vector for each class.

    A new challenge arises at this stage: this real-valued, bundled vector (which contains components with varying signs and magnitudes) must be encoded back into a quantum circuit for the final similarity comparison. The DiagonalGate operator, which we use to represent hypervectors, requires all its elements to be complex numbers with a unit modulus (absolute value of 1).

    To solve this, we developed a root-mean-square (RMS)-based phase encoding scheme. This method maps the bundled vector's components ($v_i$) to complex phases that encode both sign and relative magnitude:

- First, the RMS of the bundled vector is calculated as $RMS = \sqrt{\frac{1}{D}\sum_i v_i^2}$. This value serves as a dynamic normalization factor representing the vector's average component magnitude;
- Each component $v_i$ is then mapped to a complex phase $\phi_i = e^{i\pi v_i/RMS}$. This mapping preserves the sign (e.g., $v_i > 0 \rightarrow phase \approx 0$, $v_i < 0 \rightarrow phase \approx \pi$) while encoding the component's relative magnitude into the phase angle;
- A final DiagonalGate is constructed from the list of unit-modulus phases, creating the quantum prototype circuit.

This hybrid encoding protocol introduces a new, critical hyperparameter: the vector dimensionality ($D$). We conducted an empirical analysis to determine the optimal dimensionality for this specific task. Our results revealed a clear performance peak:

- At low dimensionality (e.g., $D = 16$, 4 qubits), the vector space is insufficiently expressive. The RMS encoding fails to capture the subtle differences between the class prototypes, resulting in a low F1 score (37.50%);
- As dimensionality increases, performance steadily rises, reaching a peak F1 score of 80.00% at $D = 128$ (7 qubits). This suggests $D = 128$ strikes an optimal balance, providing a space rich enough for the prototypes to be separable while maintaining a stable RMS normalization;
- Beyond this peak, we observed a peaking phenomenon: at $D = 256$ (8 qubits), the F1 score dropped significantly to 29.20%. This indicates that at very high dimensions, the signal in the RMS-normalized phases becomes too diluted, degrading the model's ability to classify.

Therefore, all subsequent quantum classification experiments were conducted using the empirically-determined optimal dimensionality $D = 128$.

It is also important to note that, while the full quantum bundling for this complex classification task was deemed infeasible, the LCU+OAA algorithm was successfully implemented and validated in the previous analogical reasoning task.

3. <u>Inference and classification</u>: to classify a new image from the test set, we apply the same hybrid process:
   - The test image is first encoded into its 16 classical feature vectors using the same process as in training (step 1 and 2);
   - The 16 vectors are bundled classically (step 3) to create a single query hypervector;
   - This query vector is encoded into its own single quantum circuit using a DiagonalGate;
   - The final classification is then determined by comparing the quantum state of this test sample to the quantum state of each learned class prototype;
   - This comparison is performed using the Hadamard Test. A circuit is run to compare the test sample's state with the Class 3 prototype state, and another is run to compare it with the Class 6 prototype;
   - The test sample is assigned the label of the class with which its state has the highest similarity (i.e., the highest probability of measuring $|0\rangle$ in the Hadamard Test).

A flowchart summarizing the classification task is shown in Figure 1 below.

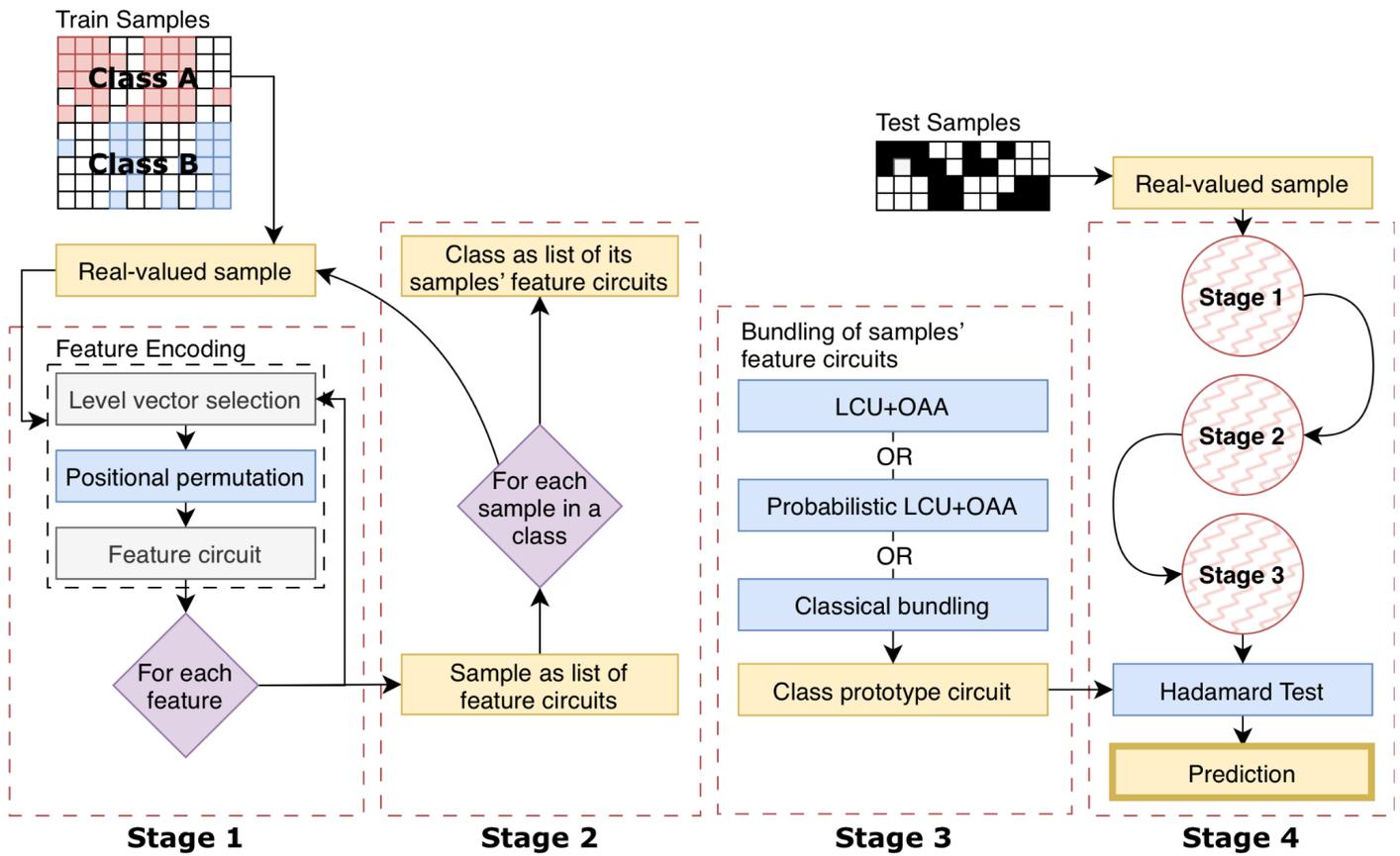

**Figure 1:** The Quantum Hyperdimensional Computing Architecture. The flowchart illustrates the end-to-end pipeline for the supervised classification task. <u>Stage 1</u> – a classical, real-valued data sample is encoded into a set of quantum feature circuits, one for each of its permuted features; <u>Stage 2</u> – to create a class prototype, the individual feature circuits from all training samples in a class are collected into a single, flat list. This entire list is then bundled exactly once using the LCU+OAA algorithm to produce the final prototype circuit; <u>Stage 3</u> – for inference, a new test sample is encoded by bundling its own feature circuits using the LCU+OAA process, a probabilistic LCU+OAA process, or the classical element-wise addition of classical vectors (then converted back into the quantum space), creating a single query circuit; <u>Stage 4</u> – the query circuit (from Stage 3) is compared against the learned prototype circuits (from Stage 2) using the Hadamard Test to determine the predicted class.

**Results and analysis** – The training and testing procedures were implemented using the hybrid bundling approach and executed on all computational platforms. Before presenting the classification results, it is crucial to visualize the core trade-off between algorithmic performance and hardware cost that dictated our experimental design.

As described in the previous section, our hybrid RMS-encoding's performance is highly dependent on the vector dimensionality. To quantify this, we ran an empirical analysis plotting the ideal noise-free simulation F1 score against the required transpiled circuit depth for the Hadamard Test at each dimensionality. The results of this analysis, which form the basis for our experimental strategy, are shown in Figure 2.

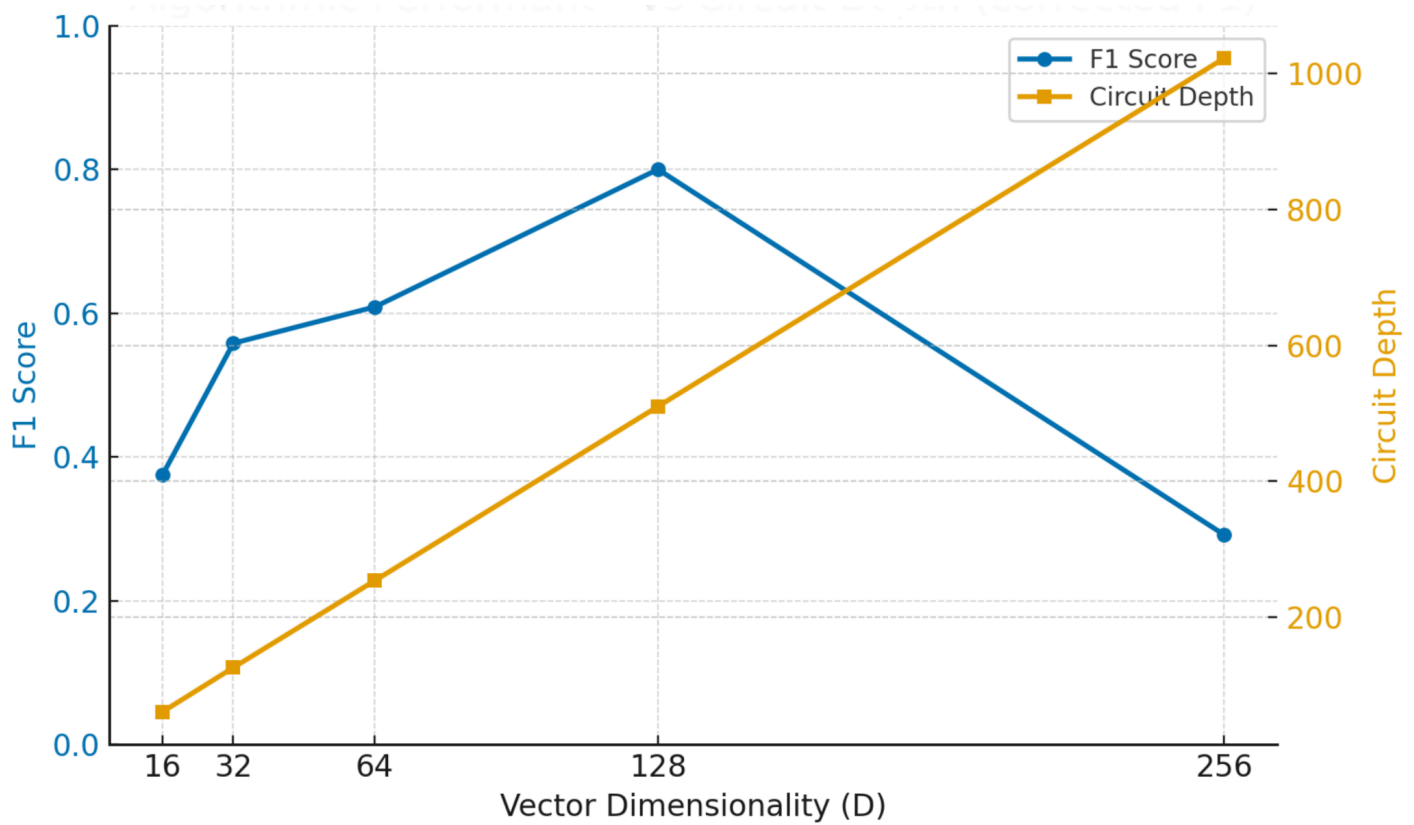

**Figure 2:** The algorithmic vs. hardware trade-off in hybrid QHDC. This plot shows the results of an empirical analysis by varying vector dimensionality ($D$). <u>Left Y-axis</u> – the ideal noise-free simulated F1 score, showing an algorithmic performance peak of 80% at $D = 128$; <u>Right Y-axis</u> – the corresponding transpiled circuit depth for the Hadamard Test, showing a rapid increase in hardware cost. This plot illustrates the peaking phenomenon of the RMS-encoding and the fundamental trade-off between algorithmic optimality ($D = 128$) and hardware feasibility ($D = 32$).

Figure 2 perfectly encapsulates the central challenge of the NISQ era:

- The algorithmic performance (F1 score) shows a clear peaking phenomenon. Performance is low at $D = 16$ (37.50%), rises sharply to an algorithmic optimum of 80.00% at $D = 128$, and then collapses again at $D = 256$ (29.16%) as the signal becomes too diluted.

- Simultaneously, the hardware cost (circuit depth) grows with dimensionality. While the depth at $D = 32$ is a manageable 126, it explodes to 510 at the $D = 128$ optimum, a depth that is practically infeasible on today's hardware.

Please note that the F1 scores shown in Figure 2 are the result of a single classification task using 100 training and 50 test samples, without any cross-validation. However, the same set of training and test samples have been used for each of the classification tasks with different vector dimensionality.

This analysis informed our two-pronged experiments that have been all performed in 5-fold cross-validation. The classification results are shown in Figure 3 ($D = 128$) and Figure 4 ($D = 32$).

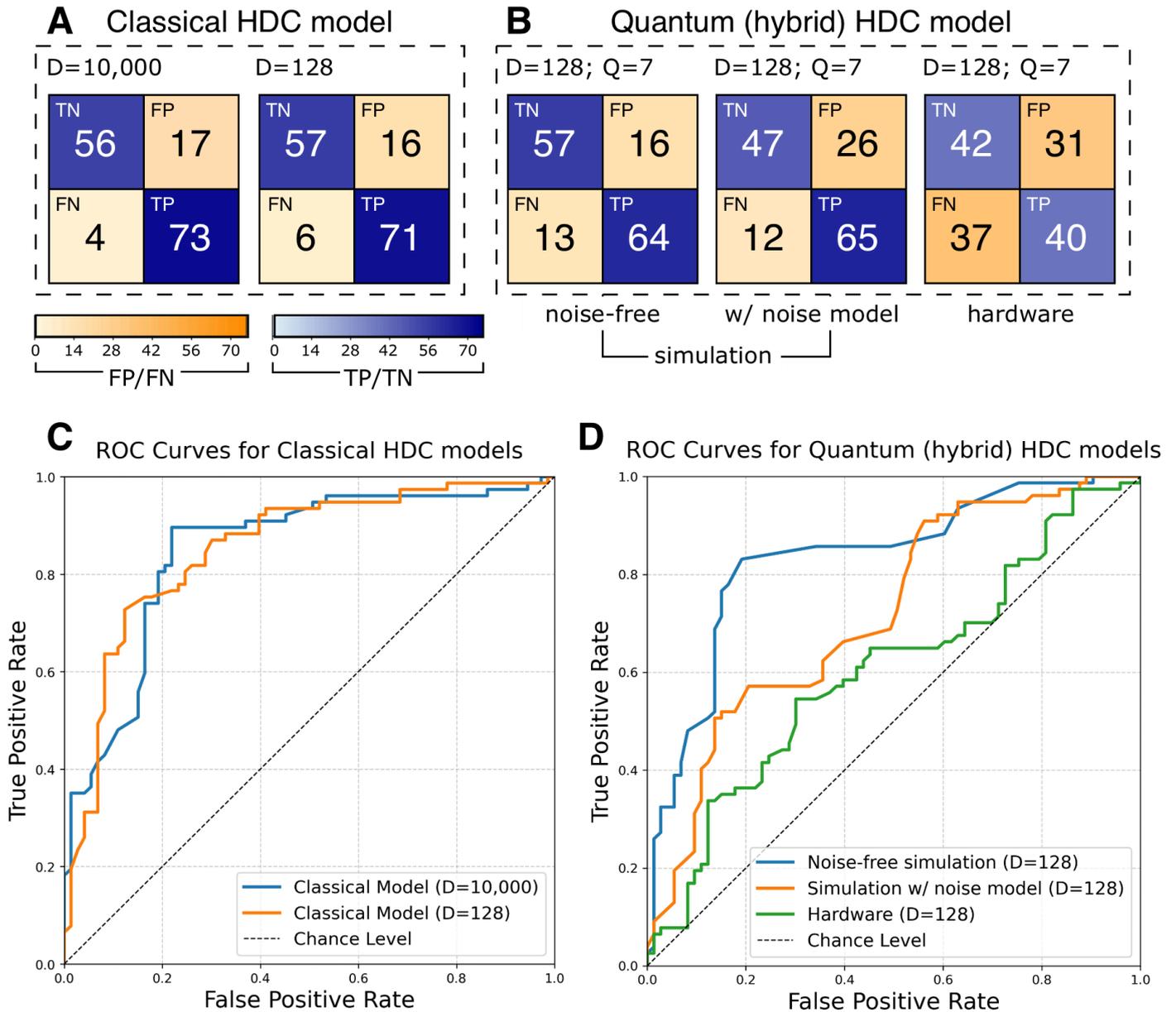

**Figure 3:** Comparative classification performance across classical, simulated, and hardware platforms. This figure displays the confusion matrices for the supervised classification task performed in 5-fold cross-validation. Each matrix shows the true class labels versus the predicted class labels, with the diagonal elements representing correct predictions. Panel A – the performance of the classical HDC model, trained and tested using conventional linear algebra using vectors with dimensionality ($D$) 10,000 and 128; Panel B – the performance of the ideal quantum model, trained and tested in a noise-free environment with quantum state vector simulations using the *AerSimulator* (left matrix), followed by the performance of the simulated quantum model configured with the noise model retrieved from *ibm_pittsburgh* (central matrix), and the performance of the quantum hardware execution, where inference for each test sample was performed on the *ibm_pittsburgh* QPU (right matrix), all using a vector dimensionality ($D$) of 128 and 7 qubits (Q). Panel C – the ROC curves of the classical HDC models ($D = 10,000$ and $D = 128$); Panel D – the ROC curves of the quantum classification experiments using a vector dimensionality ($D$) 128, visually illustrating the performance gap between the ideal noise-free, noisy, and hardware experiments.

As detailed in Figure 3 (Panel A), the classical baseline models performed robustly. The canonical model with vector dimensionality $D = 10,000$ achieved an average weighted F1-Score of 85.93% (±0.04 of standard deviation) over an average of 0.12±0.03 seconds of CPU time per cross-validation fold, while the one with vector dimensionality $D = 128$, which serves as the direct comparison for the quantum experiments, yielded an average weighted F1 score of 85.26% (±0.02 of standard deviation) over 0.05±0.00 seconds of CPU time. The slight decrease in performance is consistent with the reduced dimensionality, which offers approximately the same level of separability compared to the 10,000-dimensional space.

The quantum executions (Figure 3, Panel B), enabled by the hybrid bundling strategy and operating at the optimal $D = 128$, successfully mirrored this task. The ideal, noise-free simulation provided a theoretical benchmark with an average weighted F1 score of 80.81% (±0.04 of standard deviation) over 24.50±0.87 seconds of CPU time. When incorporating the hardware noise model, the simulation yielded an average weighted F1 score of 74.31% (±0.08 of standard deviation) over 1030.28±86.76 seconds of CPU time (approximately 17 minutes). Most importantly, the physical execution on the *ibm_pittsburgh* QPU, averaged over 5-folds, yielded a final F1 score of 54.75% (±0.09 of standard deviation) over 8527.40±38.87 seconds (approximately 142 minutes in average) of QPU usage. The sharp divergence from the noisy simulation's F1 score underscores the severe and cumulative impact of hardware noise on computations that exceed a critical circuit depth.

In Figure 3, Panel C, the ROC curves for the classical $D = 10,000$ and $D = 128$ models are nearly superimposed. This indicates near-identical behavior, with AUC values of 84.90% and 85.30% respectively. This confirms that $D = 128$ is an excellent and faithful classical proxy for the high-dimensional canonical model. On the other hand, in Panel D, the ROC curves for the simulated $D = 128$ experiments visually illustrate the impact of noise. The noise-free simulation achieves a strong AUC of 83.40%, establishing the hybrid model's theoretical viability. The noisy simulation curve shows a significant drop to an AUC of 72.00%, aligning with its 74.31% F1 score. Finally, the hardware execution curve is severely flattened, collapsing toward the diagonal line of no-discrimination and yielding a final AUC of 60.50%.

This performance drop can be directly attributed to a key trade-off between the algorithm's expressiveness and the physical constraints of NISQ hardware. Our empirical analysis identified a dimensionality of 128 (7 qubits) as the optimal dimensionality for the RMS-based hybrid encoding algorithm, yielding the highest simulated F1 score. However, this higher dimensionality creates deeper quantum circuits for the inference step. As shown in the next section, the Hadamard Test circuit required to compare a 128-dimensional sample has a transpiled depth of 510.

A circuit of this depth is highly susceptible to decoherence, which explains the significant degradation in fidelity observed on the QPU. This result suggests that while a dimensionality of 128 is algorithmically optimal for our hybrid method, its circuit depth is practically infeasible for current hardware. Conversely, a lower dimensionality $D = 32$ (5 qubits) yields a higher average weighted F1 score of 68.59% (±0.10 of standard deviation) as the effect of a much shallower Hadamard Test circuit (depth 126), with an AUC of 66.20%. Figure 4 shows the confusion matrix for the hardware execution using this hardware-aware $D = 32$ approach.

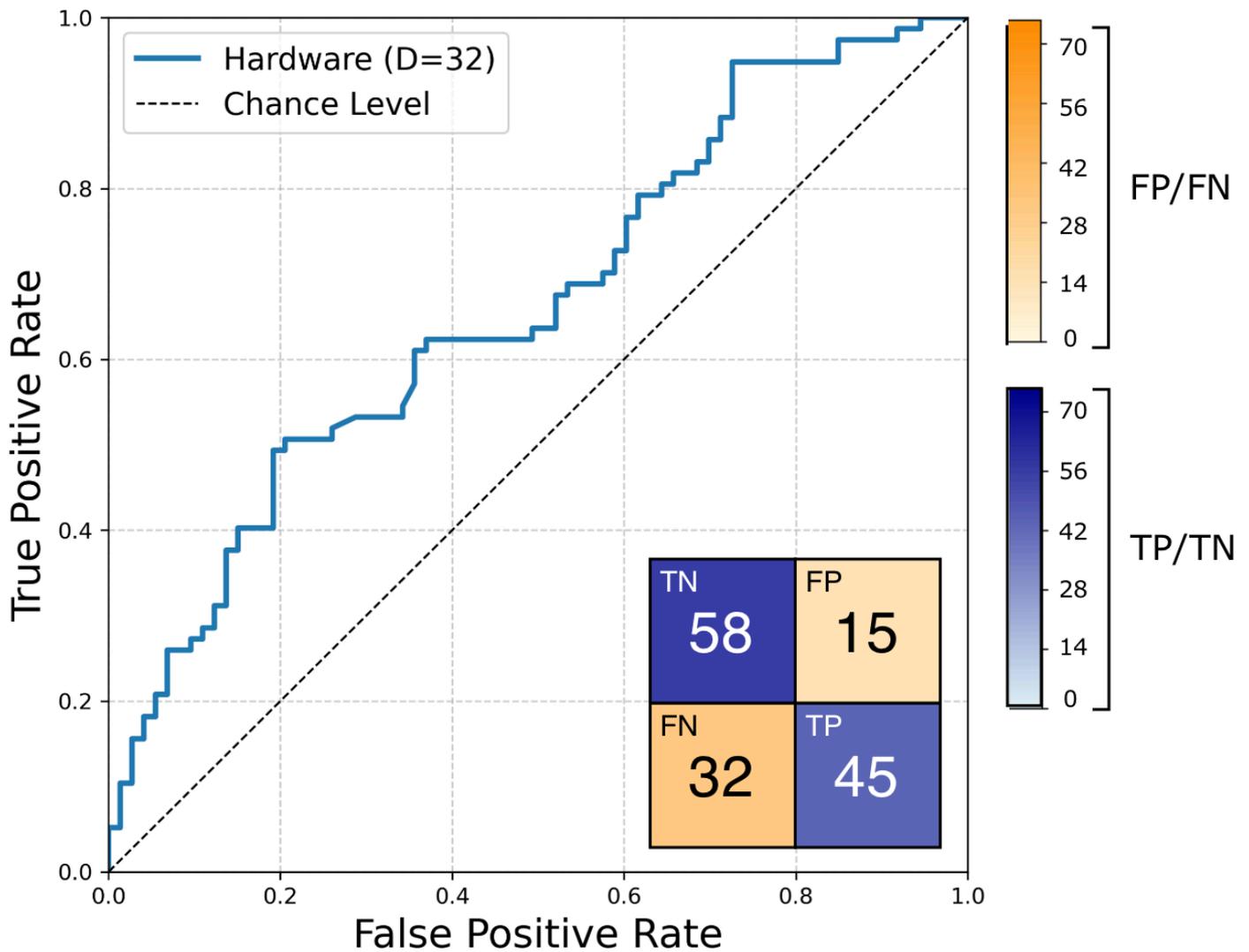

**Figure 4:** Hardware classification performance at $D = 32$. The figure displays the confusion matrix and the ROC curve for the supervised classification task, executed on the *ibm_pittsburgh* QPU using the hardware-aware dimensionality of $D = 32$ (5 qubits). The resulting shallow circuit depth (126) allowed for a significant improvement in classification fidelity compared to the $D = 128$ execution (Figure 3, Panel B, right-side matrix), validating the critical trade-off between algorithmic dimensionality and hardware fidelity.

This increase from 54.75% (at $D = 128$) to 68.59% (at $D = 32$) of average weighted F1 score over 3102.60±13.66 seconds (approximately 52 minutes in average) of QPU usage, strongly validates our hypothesis: by strategically reducing circuit depth, we successfully mitigated the worst effects of decoherence and recovered a stronger computational signal.

While the full LCU-based bundling was shown to be infeasible, these results validate that the core QHDC components are physically realizable and function correctly on NISQ hardware. This provides a robust validation for the paradigm and a clear path forward for future hardware-aware algorithm design where dimensionality is a key parameter to be tuned against hardware fidelity.

**Comparative Resource Analysis**

A central claim of this work is that QHDC possesses a near-term architectural advantage, making it well-suited for the NISQ era [46–49]. To substantiate this claim, it is essential to quantify the computational resources required by the core quantum operations developed in this framework. This section provides an analysis of the primary metrics that determine an algorithm's feasibility on current and near-term quantum hardware: circuit depth and the number of qubits entangling gates:

- Circuit depth: this metric measures the longest path of consecutive gates in the quantum circuit. In the NISQ era, circuit depth is a critical limiting factor, as longer computations are more susceptible to decoherence and cumulative gate errors. Shallower circuits are generally more likely to produce reliable results;

- CNOT count: the two-qubit CNOT gate is typically the most error-prone and time-consuming operation on many superconducting quantum computers. The total CNOT count is therefore a crucial proxy for the overall noise and complexity of a circuit.

To analyze the resource requirements, we first investigated the fully quantum-native implementation of the class prototype encoding circuit, which uses the LCU+OAA algorithm to bundle all $M \times 16$ feature circuits (where $M$ is 50 as the number of training samples in Class 3 and Class 6). The transpiled resource costs for this circuit, along with the probabilistic LCU alternative and the practical hybrid circuits, are summarized in Table 2.

| Quantum Operation | System Qubits | Ancilla Qubits | Circuit Depth | CNOT Count |
|---|---|---|---|---|
| **Full quantum-native (infeasible)** | | | | |
| Class prototype encoding (flat LCU) | 7 | 10 | 21,426,195 | 12,223,544 |
| Class prototype encoding (probabilistic LCU – 15 rounds) | 7 | 1 | 7,487 | 289 |
| **Hybrid approach (as executed)** | | | | |
| Positional permutation (single feature) | 7 | 0 | 27 | 0 |
| Hadamard Test | 7 | 1 | 510 | 0 |

**Table 2:** Quantum resource requirements for QHDC operations. The table lists the number of system and ancilla qubits, along with the transpiled circuit depth and CNOT gate count, for the principal quantum circuits used in the supervised classification experiment as resulting from the binding, bundling, and similarity estimation. It contrasts the theoretical full quantum-native LCU approaches (which are infeasible) with the practical hybrid approach that was ultimately executed on hardware.

The flat bundle LCU approach, while theoretically correct, is shown to be totally unrealizable, with a circuit depth of ~21.4 million and over 12 million CNOTs. The probabilistic LCU offers a significant improvement but it is still excessively deep for NISQ hardware at ~7,500 and 289 CNOTs, using a representative 15 random LCU rounds. This number was not fully optimized, as even this low value resulted in a circuit depth that is already prohibitively large for NISQ execution, and increasing rounds would only compound this issue.

In contrast, the hybrid bundling approach results in extremely shallow, efficient circuits for the final hardware execution. The prototype encoding (which is just a single diagonal gate) and the final Hadamard Test (which compares two diagonal gates) have depths of only 1 and 4, respectively. This demonstrates the practical necessity and viability of the hybrid approach.

While Table 2 provides a valuable snapshot, a more critical analysis involves understanding how these resource costs scale as the problem size (and thus the dimensionality of the vector space) increases. To investigate this, we performed a scaling analysis on the most resource-intensive component of our framework, the "Class prototype encoding" circuit which is the result of the LCU quantum subroutine applied on $M$ sets of 16 permuted feature circuits, with $M$ number of samples belonging to a specific class (50 in this specific example). We generated this circuit for systems with a varying number of qubits from 4 to 7 (i.e., with vector dimensionality $2^4 = 16$ to $2^7 = 128$), and measured the corresponding growth in circuit depth.

The results of this analysis, plotted in Figure 5, reveal a clear exponential growth trend. As the number of system qubits increases, the resources required by the transpiler to decompose the multi-controlled LCU operation grow rapidly.

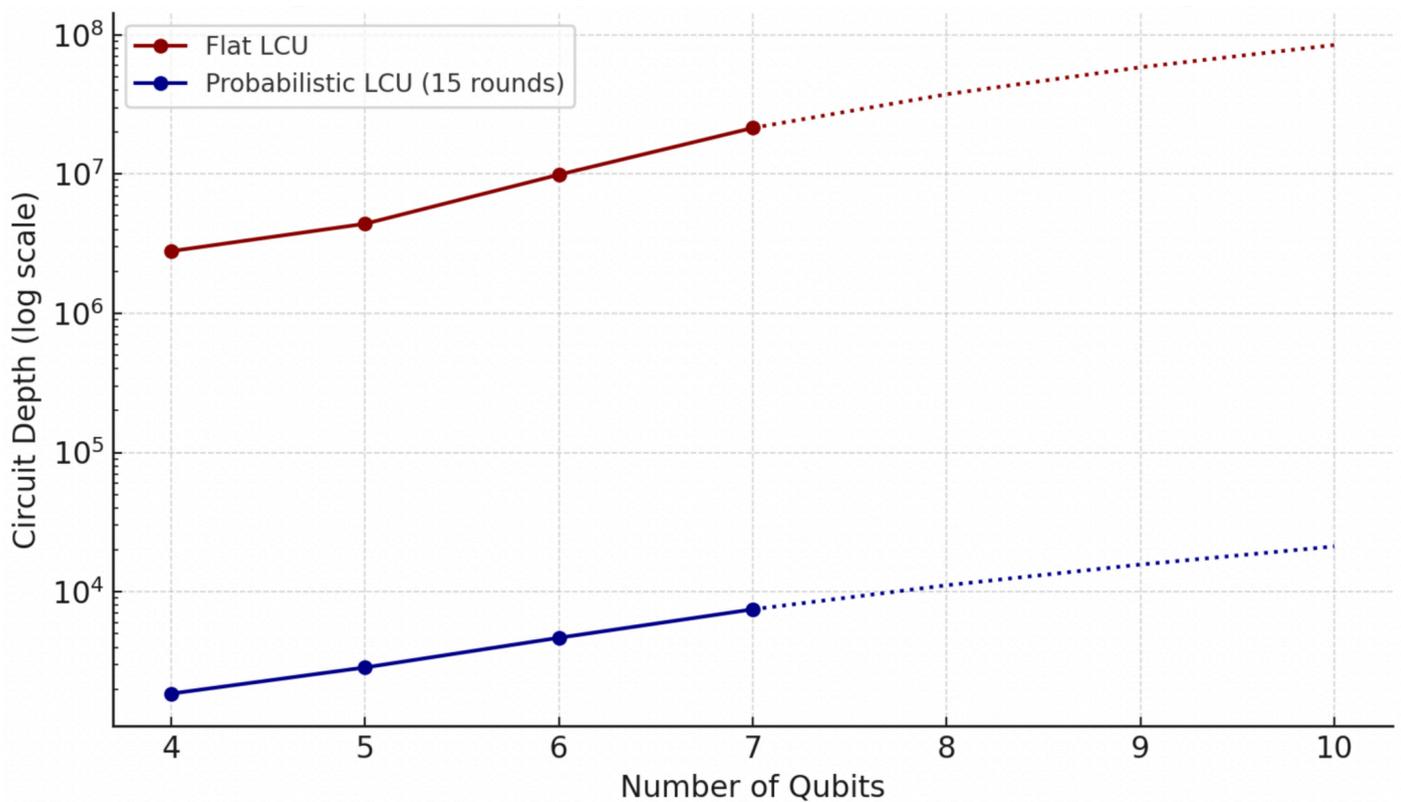

**Figure 5:** Scaling analysis of circuit depth with respect to the number of qubits for the "Class prototype encoding" circuit. The plot compares the Flat LCU (dark red) and the Probabilistic LCU (dark blue, 15 rounds) implementations. Solid lines represent measured circuit depths for 4–7 qubits, while dotted lines indicate extrapolated predictions up to 10 qubits. The logarithmic scale on the Y-axis emphasizes the exponential growth in circuit complexity for the Flat LCU and the substantially reduced depth achieved by the Probabilistic LCU approach.

Collectively, the data from Table 2 and Figure 5 reveal the primary bottleneck for scaling QHDC on NISQ devices. While the elementary binding and permutation operations are remarkably efficient, the LCU bundling algorithm's resource demands grow exponentially. This quantitative data necessitated the pivot to a hybrid bundling strategy to achieve a practical validation.

**Comparison with Established QML Paradigms**

To contextualize the performance and architectural efficiency of the QHDC framework within the broader field of QML, we conducted a direct comparison against two prominent QML paradigms: Variational Quantum Classifier (VQC) [50–53] and Quantum Support Vector Classifier (QSVC) [54–57]. As two of the most prominent and widely studied QML paradigms, VQC and QSVC serve as an essential benchmark. They were both trained and tested in the exact same synthetic dataset used in our primary classification experiment, allowing for a direct, side-by-side benchmark of not only classification accuracy but also the fundamental differences in training philosophy and the underlying quantum resource requirements. This comparison aims to highlight the unique architectural properties of QHDC relative to established iterative QML methods.

**VQC methodology** – A VQC operates on a fundamentally different principle than QHDC. It is a hybrid quantum-classical algorithm that leverages a classical computer to iteratively optimize the parameters of a quantum circuit, effectively training it to perform a specific task. Our implementation, following standard QML practices, consisted of two main quantum circuit components:

1. <u>Feature map</u>: this initial circuit is responsible for encoding classical data into the high-dimensional Hilbert space of the quantum computer. We used a standard *ZZFeatureMap* from Qiskit, a second-order Pauli feature map. This circuit applies single-qubit rotations based on the input feature values and then uses a series of entangling CNOT and controlled-Z gates. The purpose of this entanglement is to create complex, non-linear correlations between the features in the quantum state space;

2. <u>Variational ansatz</u>: following the feature map, a trainable circuit known as the ansatz is applied. We used a *RealAmplitudes* circuit, a common choice for VQCs. This circuit consists of alternating layers of single-qubit rotational gates and entangling CNOT gates. The angles of the rotational gates are the trainable parameters of the model, analogous to the weights in a classical neural network.

The training process involves an iterative optimization loop that alternates between the quantum and classical processors. For each training sample, the full quantum circuit is executed, and the resulting state is measured to produce a prediction. A classical cost function then evaluates the difference between this prediction and the true label. Based on this cost, a classical optimization algorithm suggests new values for the rotation angles in the ansatz. This process is repeated for many iterations until the model's accuracy on the training data converges.

Note that, we used 1 number of repeated circuits for defining the feature map as a foundational choice to test the model with a direct, non-complex feature embedding. On the other hand, we chose 3 repeated circuits (layers) for the variational ansatz to provide a sufficient number of trainable parameters for the model to capture non-trivial relationships in the data. We also used a maximum of 100 iterations for the COBYLA optimizer [58,59] to balance the search for an optimal solution against the computational cost of the iterative training loop within the cross-validation.

**QSVC methodology** – A QSVC operates on a fundamentally different principle from both QHDC and VQC. It is a hybrid quantum-classical algorithm, but instead of training a parameterized circuit, it uses a quantum computer to map data into a high-dimensional feature space and compute a quantum kernel (i.e., a similarity measure between data points within that space). This kernel is then consumed by a classical Support Vector Machine (SVM) algorithm. Our implementation, following standard QML practices, was built around Qiskit's *FidelityQuantumKernel* and consisted of two main components:

1. <u>Feature map</u>: this circuit is responsible for encoding classical data into the high-dimensional Hilbert space, which defines the feature space where the kernel is computed. Again, we used a standard *ZZFeatureMap*. This circuit applies single-qubit rotations based on input features and entangling gates. Its purpose is to create complex, non-linear correlations between the features, projecting the data into a space where it is hopefully more easily separable;

2. <u>Quantum kernel and fidelity</u>: the second key component is the kernel calculation itself. Here, we used the *FidelityQuantumKernel*. This object takes the feature map and a fidelity-measurement routine as input. For the fidelity measurement, we used *ComputeUncompute*, a method that estimates the similarity between two quantum states, $\left|\langle\phi(x_i)|\phi(x_j)\rangle\right|^2$ which forms the entries of the kernel matrix.

The training process involves the classical SVM algorithm iteratively querying the quantum computer. To build the kernel matrix, the *FidelityQuantumKernel* must be evaluated for pairs of data points on the quantum simulator (using the *ComputeUncompute* method). Once this kernel matrix is constructed, the classical SVM algorithm computes the optimal separating hyperplane in the quantum feature space. The quantum circuit parameters themselves are fixed and are not optimized during this process.

Please note that, for the *ZZFeatureMap*, we used 1 repetition of the circuit in this case as well as a foundational choice to define the feature space. We also adopted a linear entanglement as a specific structural choice for the feature map, creating entangling gates only between adjacent qubits. This defines the specific class of non-linear correlations to be explored by the classical SVM.

**Comparative analysis** – the comparison of QHDC, VQC, and QSVC frameworks was based on three key aspects: the training paradigm, the resource required for a single inference operation, and the final classification performance:

● <u>Training paradigm</u>: a profound architectural difference lies in how each model learns. The VQC and QSVC both rely on iterative, optimization-based training. The VQC performs a gradient-descent-like search (though COBYLA is gradient-free) to find optimal parameters, while the QSVC must repeatedly query the QPU to construct its kernel matrix. Both approaches can be computationally intensive. In contrast, the hybrid QHDC model learns in a single, deterministic one-shot process by bundling the classical vectors, completely avoiding the iterative optimization loops and potential challenges like barren plateaus or convergence to local minima, or other challenges associated with variational training;

● <u>Resource efficiency (inference)</u>:  the operational cost of using a trained model is determined by the resources required for a single inference step. This involves preparing a quantum state for a new data point and performing a measurement or comparison. In terms of resource cost, we see a potential advantage of the QHDC architecture. The QHDC sample encoder circuit, while deep, is constructed based on the algebraic properties of the data. VQC circuits, particularly with complex feature maps and layered ansatz, can often result in deeper circuits with higher CNOT counts for a comparable number of qubits. On the other hand, QSVC requires an inference cost that scales with the number of support vectors identified during training, as the quantum kernel must be evaluated between the new data point and each of these vectors to make a final decision;

● <u>Performance</u>: ultimately, the efficacy of any classifier is measured by its ability to generalize to unseen data. The performance of VQC and QSVC was evaluated on its ability to correctly classify the samples in a 5-fold cross-validation scenario. The results in terms of the final confusion matrices and ROC curves, are presented in Figure 6, allowing for a direct comparison against the QHDC performance results previously shown in Figure 3.

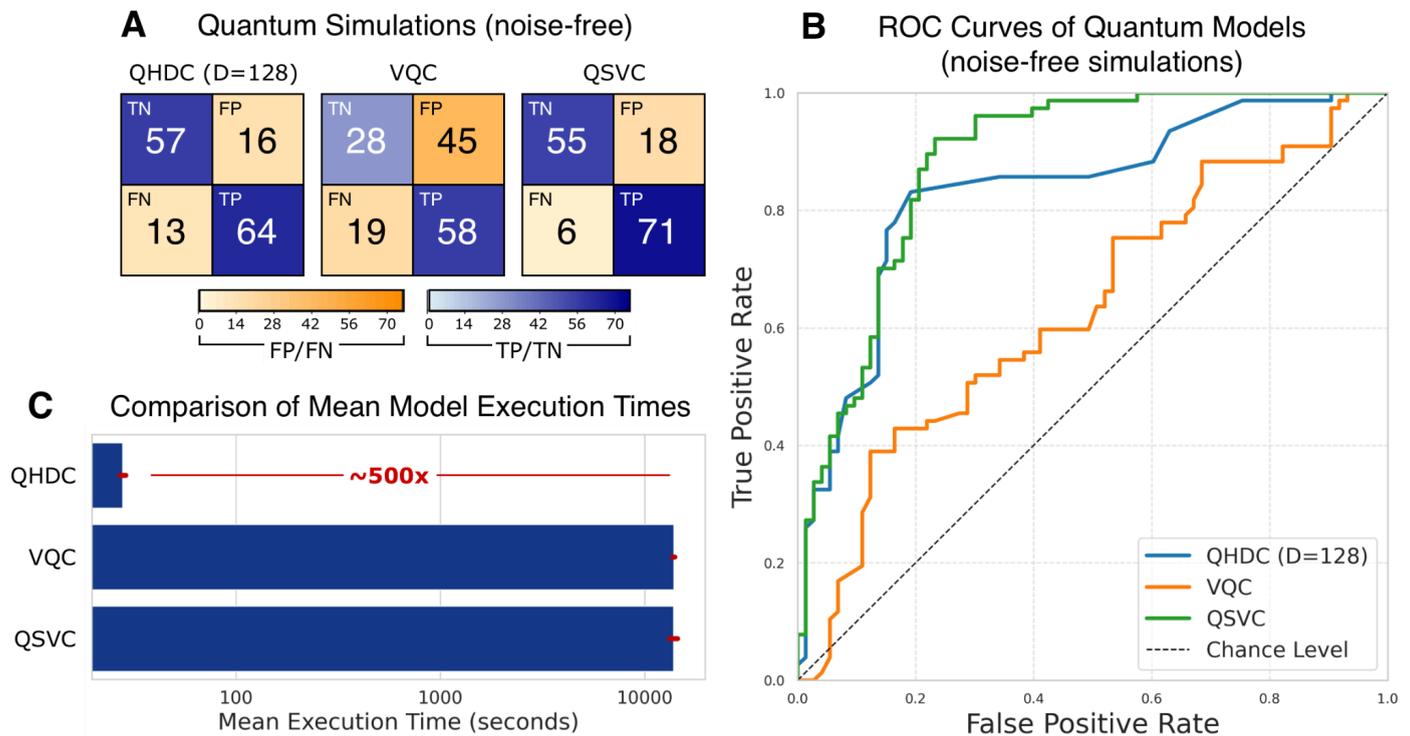

**Figure 6:** Comparative benchmark of QHDC ($D = 128$), VQC, and QSVC, all executed in an ideal, noise-free simulation. <u>Panel A</u> – confusion matrices showing the true class labels against the predicted class labels, with the diagonal elements representing correct classifications; <u>Panel B</u> – ROC curves for both VQC and QSVC on the unseen test test (aggregated from 5-fold cross-validation); <u>Panel C</u> – comparison of the mean execution time (in seconds) required to complete a single cross-validation. The X-axis is on a logarithmic scale to visualize the vast difference in magnitude between QHDC and the other models, with a speedup of ~500x.

The most profound difference is in computational cost, as illustrated in Figure 6 (Panel C). The iterative, optimization-based training paradigm of the VQC and QSVC are computationally expensive, requiring an average of 13,853.10±161.69 seconds and 13,838.22±535.87 seconds per cross-validation fold respectively (~3.8 hours each). In sharp contrast, the QHDC ($D = 128$) model requires only 27.90±0.80 seconds on average. This represents a computational speedup of ~500x, demonstrating a significant architectural advantage by avoiding iterative optimization loops entirely.

In terms of classification efficacy (Figure 6, Panels A and B), the QHDC ($D = 128$) model achieved a robust average weighted F1 score of 80.81% (±0.04 of standard deviation) and an AUC of 83.40% (as previously shown in Figure 3). The QSVC model, leveraging its quantum kernel, performed slightly better than QHDC, yielding an average weighted F1 score of 84.01% (±0.03 of standard deviation) and an AUC of 84.01%. The VQC model, however, struggled to converge to a strong solution in this experiment, finishing with the lowest average weighted F1 score of 54.46% (±0.11 of standard deviation) and a corresponding AUC of 69.75%.

Collectively, these results demonstrate the QHDC model achieves highly competitive classification performance, while also being orders of magnitude more computationally efficient in terms of run time required to train and evaluate the classification model compared to state-of-the-art QML methods.

**DISCUSSION AND CONCLUSIONS**

In this work, we have introduced and validated Quantum Hyperdimensional Computing (QHDC), a novel quantum neuromorphic paradigm that operates natively on quantum hardware. We established a formal mapping between the core operations of Hyperdimensional Computing (HDC) and the fundamental principles of quantum mechanics. Specifically, we demonstrated that (i) the binding operation maps elegantly onto quantum phase oracles, (ii) the bundling operation can be realized through sophisticated quantum subroutines like the LCU and OAA, and (iii) the permutation operation can be performed with a QFT.

We move this framework from theory to practice, first of its kind, by presenting the first physical validation on a quantum computer. Through two distinct experiments (i.e., a symbolic analogical reasoning task and a data-driven supervised classification task) we show the framework's versatility. In this sense, this work is the first to (i) jointly establish the theoretical foundations of QC + HDC's complementary process with realistic hardware performance proof, (ii) realize a full software implementation, and (iii) validate the approach on real quantum hardware across different application domains both for *reasoning* and *classification*. The viability of this approach was rigorously assessed through a comparative analysis across three computational platforms: a classical baseline, an ideal noise-free quantum simulation, and execution on a state-of-art IBM Heron r3 QPU. The results provide compelling evidence that QHDC is a physically realizable and computationally coherent paradigm. Ultimately, this work not only validates this but also opens the door to a new class of neuromorphic-inspired quantum machine learning (QML) algorithms.

**Notes on Quantum Utility and Advantage**

A critical question for any new quantum algorithm is its potential benefit over classical methods. For QHDC, it is useful to distinguish between *quantum utility*, which refers to a quantum approach providing a valuable, near-term pathway to solving a problem, and *quantum advantage*, the long-term goal of demonstrably outperforming the best classical algorithms. We see a clear path from QHDC's current utility to its future advantage.

**Near-Term Quantum Utility** – The primary benefit of QHDC in the current Noisy Intermediate-Scale Quantum (NISQ) era, where gate fidelity and qubit coherence are limited resources, is one of quantum utility, rooted in its architecture. Unlike classical algorithms ported to a quantum machine, QHDC is quantum-native. The one-to-one mapping of HDC operations onto core quantum subroutines means that it can, in principle, utilize quantum hardware more efficiently and with shallower circuits than many existing QML models. Our implementation demonstrates this utility in practice: by avoiding the iterative optimization loops of benchmark QML models like VQC and QSVC, our QHDC model achieved a computational speedup of nearly 500x in simulation.

**Long-Term Quantum Advantage** – The long-term vision for QHDC is to achieve true quantum advantage as hardware matures. The exponential capacity for qubits to represent information, combined with quantum parallelism, could offer significant, non-classical speedups in several key areas. For instance, large-scale similarity searches, such as finding the closest match for a patient's genomic profile in a database of millions, could potentially be performed in a single quantum operation, leveraging algorithms analogous to Grover's search [60,61]. Furthermore, the exploration of complex, structured relationships, a hallmark of HDC's reasoning capabilities, could be massively accelerated by creating deeply entangled states across multiple quantum registers simultaneously. Realizing this advantage will require the advent of fault-tolerant quantum computers, but the quantum-native structure of QHDC provides a direct and promising roadmap toward that goal.

**Long-Term HDC Advantage** – Our findings highlight not only how HDC can be implemented on quantum hardware, but also how quantum processes can inform the design of classical HDC systems. While long hypervectors are traditionally preferred in HDC, a key challenge is to reduce dimensionality without sacrificing accuracy. Our results reveal that, when combined with complementary quantum operations, short-dimensional hypervectors can still preserve robust performance, suggesting that hundreds of dimensions may suffice where thousands were once required. This opens the door to future work in which offline quantum procedures can be used to "handcraft" hypervector designs for classical HDC, potentially yielding more orthogonal, hardware-aware vector sets than those obtained through purely classical random initialization.

**Current Limitations**

While this work establishes a successful validation of our paradigm, it is essential to acknowledge the practical limitations that must be addressed for QHDC to scale effectively in the NISQ era and beyond. These challenges represent key avenues for future research.

The most significant limitation is the explosion of circuit depth and gate count inherent in the LCU+OAA bundling algorithm. In our model, the class prototype is constructed from a single LCU operation over a large list containing the feature circuits from all training samples. The complexity of this LCU operation scales directly with the total number of unitaries being bundled. As the training set grows, this list becomes very large (e.g., $M$ samples $\times$ 16 features), causing the transpiler to decompose the multi-controlled operations into a vast number of primitive gates. This results in a final prototype circuit whose depth is too great to be executed with high fidelity on current quantum hardware.

Our investigation into a probabilistic LCU variant, which uses a single ancilla to control one unitary at a time over a set number of rounds, offered a significant optimization. This approach reduced the circuit depth drastically by orders of magnitudes. However, this depth, while a massive improvement and feasible in simulation, is still too great for high-fidelity execution on today's noisy quantum hardware.

This finding necessitated our pivot to a hybrid quantum-classical bundling strategy for the classification experiment. By strategically moving the bundling operation to the classical domain by retrieving the classical vectors from their quantum circuit definitions, we were able to isolate this bottleneck. This practical compromise was essential, as it allowed us to successfully validate the other core components of the QHDC framework.

The adoption of a full quantum approach for quantum neuromorphic classification systems will require research into more advanced circuit synthesis and transpilation techniques [62–64] specifically designed for LCU-based algorithms, which could potentially find more efficient decompositions.

Furthermore, classical HDC models often employ an iterative retraining process to refine class prototypes by adding correctly classified samples and subtracting misclassified ones [29,65–70]. Implementing such a feedback loop in QHDC presents both a challenge and an exciting opportunity. The challenge lies in the immense cost of each iteration, which would require reconstructing and re-running the entire deep LCU+OAA bundling circuit with the new list of feature circuits. However, this process could also function as a novel form of algorithmic error mitigation. The inherent noise-tolerance of classical HDC suggests that iteratively reinforcing the prototype could mitigate the effects of hardware noise, essentially allowing the algorithm's structure to average out physical errors over time. Investigating whether the corrective benefits of this retraining can outweigh the detrimental effects of noise compounding from repeated deep-circuit executions is a compelling direction for future research.

Additionally, like all quantum algorithms, the performance of QHDC on physical hardware will be constrained by noise and qubit connectivity. The LCU and OAA algorithms rely on precise phase interference, which is highly susceptible to decoherence and gate errors. Future work should involve a detailed analysis of the framework's resilience to different types of quantum noise and the development of tailored error mitigation strategies to improve its performance on NISQ devices [71–73].

Finally, a critical practical limitation that underpins all NISQ-era research is the immense resource cost, in both time and monetary expense, of hardware execution. As demonstrated in our classification experiment (which required hours of QPU time), running the deep or numerous circuits required for validation is prohibitively expensive. This reality provides a powerful economic and practical incentive for future research to focus relentlessly on algorithmic optimization. This includes not only developing more efficient algorithms for performing bundling, but also creating novel, ultra-shallow circuit designs for the complete QHDC arithmetic.

**Future Perspectives: potential applications of quantum neuromorphic computing**

The successful implementation of the QHDC framework on quantum hardware opens a promising frontier for tackling problems whose complexity is rooted in high-dimensional, structured data. While the full potential of this paradigm will be realized with the maturation of fault-tolerant quantum computers, we can already project several high-impact application areas where QHDC could offer a transformative advantage over classical systems, especially within biomedical sciences [20,21]. Indeed, the successful application of these brain-inspired models, even on classical hardware, depends critically on robust design choices to avoid common pitfalls [74]. The QHDC framework inherits these design principles while offering a new, powerful substrate for their execution.

**Large-scale genomics and proteomics analysis** – Modern biology is flooded by new sequencing data. Searching massive databases for sequences similar to a query is a fundamental task. Current state-of-the-art tools are indispensable but mostly rely on heuristics and seed-based methods to manage the immense computational cost, meaning they are not guaranteed to find the optimal alignment.

HDC is naturally suited for sequence analysis. A DNA, RNA, or protein sequence can be encoded into a single hypervector via binding its n-grams to their positions [70,75–77]. Because genomics applications involve a small, fixed alphabet rather than an overwhelming number of features that would complicate the QHDC architecture, they are an excellent fit. The resulting hypervector represents the entire sequence holistically, and the similarity between two sequences can be computed with a single dot product, yielding a robust, alignment-free method for similarity search.

With QHDC, a database containing millions of genomic sequences could be loaded into a single, massive quantum state via superposition. A query sequence, also encoded as a quantum state, could then be compared against the entire database simultaneously. Using quantum algorithms analogous to Grover's search, one could identify the best match with a potential quadratic speed. This would enable, for the first time, exact, full-database similarity searches, moving beyond the limitations of classical heuristics.

**Accelerated drug discovery and cheminformatics** – The initial phase of drug discovery involves screening vast virtual libraries, often containing billions of molecules, to find candidates that bind effectively to a target protein. This virtual screening is a computationally expensive task, limited by the speed at which classical computers can calculate the interaction between a candidate molecule and its target.

The compositional nature of HDC is perfect for representing molecules. A molecule can be encoded as a hypervector by bundling the bound representations of its constituent atoms, functional groups, and their relative positions [78,79].

As with genomics, an entire library of drug candidates could be encoded into a single superposition state on a quantum computer. A query vector, representing the key features of a protein's binding site, could then be compared against all molecules in the library simultaneously using the Hadamard Test. This massive parallelism could drastically accelerate the initial screening phase, allowing researchers to explore a much larger and more complex chemical space than is currently feasible, significantly increasing the probability of discovering novel therapeutic agents.

**Multimodal data fusion for personalized medicine** – One of the most significant challenges in modern artificial intelligence is the fusion of multimodal data, i.e., the ability to integrate and reason over information from disparate sources, such as images, text, time-series, and tabular data. Conventional machine learning approaches often struggle with this task, typically requiring separate, specialized models for each modality that are difficult to combine into a single, coherent representation [80–82].

HDC offers a powerful and elegant solution to this problem. Because all data, regardless of its original form, is encoded into the same high-dimensional vector space, HDC provides a unified mathematical framework for representation. Information from any modality can be mapped to a hypervector [83]. For example, a clinical diagnosis from text, a genomic sequence, and an MRI scan can all coexist as points in the same space. Using the bind and bundle operations, these representations can be combined into a single, holistic hypervector that captures the complete state of a complex entity.

This capability is particularly compelling in the context of personalized medicine, a field whose promise rests on the ability to integrate vastly different types of patient data, from multi-omics data (including genomics, transcriptomics, proteomics, metabolomics, etc.) to clinical notes, lab results, and medical imaging [84]. By leveraging the principles of HDC, a patient's entire medical history can be encoded into a single patient hypervector. This creates a holistic, queryable representation of the patient's state that is far richer than any single data point.

With QHDC, it becomes possible to perform complex analogical reasoning across an entire patient population. For example, one could ask "*which patients in our database have a genomic profile similar to Patient A and a response to Treatment B similar to Patient C?*". This complex query could be performed efficiently on a quantum computer by preparing the query as a quantum state and comparing it against a superposition of the entire patient database.

These examples, while diverse, share a common theme: they are all fundamentally limited by the classical challenge of searching, comparing, and reasoning with vast amounts of structured and unstructured information. The QHDC framework, with its ability to represent complex entities as single quantum states and leverage quantum parallelism, offers a new and powerful set of tools for addressing these foundational problems. While we have focused on applications in biomedical sciences, the same principles could readily be applied to other data-intensive domains such as materials science, financial modeling, and climate science, among other scientific fields. Ultimately, this work represents the first steps toward a new era of quantum-native computation for scientific discovery across all disciplines.

**ADDITIONAL INFORMATION**

**Abbreviations**

AUC – Area Under the Curve
COBYLA – Constrained Optimization By Linear Approximation
HDC – Hyperdimensional Computing
LCU – Linear Combination of Unitaries
MAP – Multiply-Add-Permute
NISQ – Noisy Intermediate-Scale Quantum
OAA – Oblivious Amplitude Amplification
QHDC – Quantum Hyperdimensional Computing
QC – Quantum Computer
QFT – Quantum Fourier Transform
QML – Quantum Machine Learning
QPU – Quantum Processing Unit
ROC – Receiver Operating Characteristic
SVM – Support Vector Machine
VQC – Variational Quantum Classifier
VSA – Vector-Symbolic Architectures

**Availability**

The *hdlib* 2.1 Python package is distributed through the Python Package Index (*pip install hdlib>=2.1.0*) and the Conda package manager (*conda install hdlib>=2.1.0 -c conda-forge*), while its code is open-source and it is available in GitHub at https://github.com/cumbof/hdlib under the MIT license.

The quantum implementations presented here rely on IBM's Qiskit framework. The necessary packages, *qiskit* (>=2.2.1), *qiskit-aer* (>=0.17.2), *qiskit-ibm-runtime* (>=0.42.0), and *mthree* (>=3.0.0) are also publicly available through the Python Package Index (*pip install "qiskit>=2.2.1" "qiskit-aer>=0.17.2" "qiskit-ibm-runtime>=0.42.0" "mthree>=3.0.0"*).

The complete source code for the analogical reasoning and supervised classification experiments presented here, including implementations for classical, simulated quantum, and quantum hardware execution, is provided within the *hdlib* package's examples on GitHub at https://github.com/cumbof/hdlib/tree/main/examples ensuring full reproducibility.

**Author Contribution**

FC conceived the research; FC, RL, BR, and JJ discussed the theoretical framework; FC developed the mapping with HDC arithmetic; FC implemented the framework and designed the quantum circuits; FC and BR performed and discussed the comparative analysis of QHDC against state-of-the-art QML approaches; FC performed the classical, simulated, and quantum hardware experiments; FC, BR, JJ, AM, SA, and DB analyzed and discussed the results; DB supervised the research; all authors read, revised, and approved the final manuscript.

**Conflict of Interests**

Authors have no conflicts to disclose.

**Acknowledgments**

The authors would like to acknowledge the use of a Large Language Model (Google's Gemini 2.5 Pro) for its assistance in rephrasing sentences and improving the overall clarity of the manuscript's prose. All scientific ideas, mathematical formulations, the algorithmic framework, and software implementations presented in this work are the exclusive product of the authors' experience and expertise.